\documentclass[a4paper,11pt]{article}

\usepackage[a4paper,left=80pt,right=80pt,top=80pt,bottom=105pt]{geometry}
\usepackage[latin1]{inputenc}
\usepackage{amsfonts,amsmath,amssymb}
\usepackage{amsthm}
\usepackage{graphicx}
\usepackage[small,bf]{caption}
\usepackage{subcaption}
\usepackage{axodraw}
\usepackage{booktabs}
\usepackage[numbers,sort]{natbib}
\usepackage{color}
\usepackage{ulem}
\usepackage{tikz}
\usepackage{multirow}
\usepackage[pagebackref=false]{hyperref}

\usetikzlibrary{arrows}
\allowdisplaybreaks[1]

\def\bal#1\eal{\begin{align}#1\end{align}}
\newcommand{\be}{\begin{equation}}
\newcommand{\ee}{\end{equation}}
\newcommand{\besub}{\begin{subequations}}
\newcommand{\eesub}{\end{subequations}}
\newcommand{\ba}{\begin{array}}
\newcommand{\ea}{\end{array}}
\newcommand{\bi}{\begin{itemize}}
\newcommand{\ei}{\end{itemize}}
\newcommand{\nn}{\nonumber}

\newcommand{\eq}[1]{Eq.~(\ref{#1})}

\newcommand{\vev}[1]{\ensuremath{\langle #1 \rangle}}
\newcommand{\eV}{{\rm eV}}
\newcommand{\GeV}{{\rm GeV}}
\newcommand{\TeV}{{\rm TeV}}

\newcommand{\ckm}{{\text{\tiny CKM}}}
\newcommand{\mns}{{\text{\tiny PMNS}}}

\def\nl{.}
\def\nlone{.}

\begin{document}

\begin{titlepage}

\vspace*{-15mm}
\vspace*{0.7cm}

\begin{center}
{\huge A flavour GUT model with $\theta_{13}^{\raisebox{4pt}{\textrm{\scriptsize PMNS}}} \simeq\, \theta_C/\sqrt{2}$}
\\[8mm]

Stefan Antusch$^{\star\dagger}$\footnote{Email: \texttt{stefan.antusch@unibas.ch}},~
Christian Gross$^{\star}$\footnote{Email: \texttt{christian.gross@unibas.ch}},~
Vinzenz Maurer$^{\star}$\footnote{Email: \texttt{vinzenz.maurer@unibas.ch}},~
Constantin Sluka$^{\star}$\footnote{Email: \texttt{constantin.sluka@unibas.ch}}
\end{center}
\addtocounter{footnote}{-4}

\vspace*{0.20cm}

\centerline{$^{\star}$ \it
Department of Physics, University of Basel,}
\centerline{\it
Klingelbergstr.\ 82, CH-4056 Basel, Switzerland}

\vspace*{0.4cm}

\centerline{$^{\dagger}$ \it
Max-Planck-Institut f\"ur Physik (Werner-Heisenberg-Institut),}
\centerline{\it
F\"ohringer Ring 6, D-80805 M\"unchen, Germany}

\vspace*{1.2cm}

\begin{abstract}
\noindent
We propose a supersymmetric SU(5) GUT model with an $A_4$ family symmetry -- including a full flavon- and messenger sector -- which realises the relation $\theta_{13}^\mns \simeq  \theta_C / \sqrt{2}$. The neutrino sector features tri-bimaximal mixing, and $\theta_{13}^\mns \simeq  \theta_C / \sqrt{2}$ emerges from the charged lepton contribution to the PMNS matrix, which in turn is linked to quark mixing via specific GUT relations. These GUT relations arise after GUT symmetry breaking from a novel combination of group theoretical Clebsch-Gordan factors, which in addition to large $\theta_{13}^\mns$ lead to promising quark lepton mass ratios for all generations of quarks and leptons and to $m_s/m_d =18.95_{-0.24}^{+0.33},$ in excellent agreement with experimental results. The model also features spontaneous CP violation, with all quark and lepton CP phases determined from family symmetry breaking. We perform a full Markov Chain Monte Carlo fit to the available quark and lepton data, and discuss how the model can be tested by present and future experiments.
\end{abstract}

\end{titlepage}
\newpage

\section{Introduction}

The longstanding supposition that the mixing angle $\theta_{13}^\mns$ of the Pontecorvo-Maki-Nakagawa-Sakata (PMNS) matrix $U_\mns$ is possibly not only small but actually vanishing has recently been undone by the data from T2K~\cite{Abe:2011sj}, Double Chooz~\cite{Abe:2011fz}, RENO~\cite{Ahn:2012nd}, and in particular Daya Bay~\cite{An:2012eh}.
A global fit by the NuFIT collaboration~\cite{GonzalezGarcia:2012sz} finds $\theta_{13}^\mns={8.75^\circ} _{-0.44^\circ}^{+0.42^\circ} $.
This has stimulated a significant amount of interest and excitement in the neutrino flavour physics community.

While one might say that the fact that the reactor angle is not very small does not strengthen the viewpoint that the neutrino masses and mixings are determined by an underlying organising principle, there nevertheless are many interesting open roads for obtaining the neutrino parameters in flavour models.
In particular, one may, as e.g.~in tri-bimaximal (TBM) mixing models~\cite{Harrison:2002er}, adhere to the assumption that the 1-3 mixing in the neutrino sector, $\theta_{13}^{\nu}$, is vanishing and that the measured nonzero value of $\theta_{13}^\mns$ arises from the charged lepton sector only.

This case is especially interesting in the framework of Grand Unified Theories (GUTs) where the Yukawa matrices for the charged leptons and for the down-type quarks have the same origin.
One may in that case be tempted to ask whether the fact that $\theta_{13}^\mns$ agrees well with $\theta_C / \sqrt{2} \simeq 9.2^\circ$ (where $\theta_C$ is the Cabibbo angle) is more than a mere coincidence.
While it was proposed already many years ago that $\theta_{13}^\mns$ could/should be of the {\it order} of the Cabibbo angle (see e.g.~\cite{cabbibo}), the possibility that the specific relation $\theta_{13}^\mns\simeq\theta_C / \sqrt{2}$ (up to subleading corrections) emerges out of a realistic GUT has been discussed just recently~\cite{Antusch:2012fb}.\footnote{As a phenomenological possibility, the relation $\theta_{13}^\mns = \theta_C / \sqrt{2}$ was mentioned already some time ago~\cite{Minakata:2004xt}. Its possible origin from charged lepton corrections in Pati-Salam models has been discussed in~\cite{Antusch:2011qg,King:2012vj}. In SU(5) GUTs, predictions for large $\theta_{13}^\mns$ from charged lepton corrections with consistent quark-lepton mass relations were studied in~\cite{Antusch:2011qg,Marzocca:2011dh}, and conditions for realising $\theta_{13}^\mns = \theta_C / \sqrt{2}$ were given in~\cite{Antusch:2012fb}.}

In Ref.~\cite{Antusch:2012fb}, four simple conditions on a flavour GUT model were shown to be sufficient to obtain $\theta_{13}^\mns\simeq\theta_C / \sqrt{2}$: 
(i) $\theta_{13}^\mns$ should arise solely from the 1-2 (and not 1-3) mixing in the charged lepton sector, (ii) the 1-2 mixing in the down quark sector should approximately equal the Cabibbo angle, (iii) the relevant entries in the charged lepton and down quark Yukawa matrices should be generated by a single GUT-operator, and, (iv) the relevant GUT operators should feature certain ratios of Clebsch-Gordan factors (which appear after GUT symmetry breaking in the charged lepton Yukawa matrix).
Moreover, it was discussed that~-- in the case of a supersymmetric (SUSY) SU($5$) GUT with vanishing 1-1 element in $Y_d$ and $Y_e$ -- only a single combination of the Clebsch factors discussed in~\cite{Antusch:2011qg} is viable, namely $c_{12}=6$, $c_{21}=-1/2$ and $c_{22}=6$ (where $c_{ij}$ is the Clebsch of the i-j element of $Y_e$).\footnote{Following a different approach, an SU(5) $\times$ T' model with large $\theta_{13}^\mns$ close to the current experimental best fit value has been constructed in~\cite{Meroni:2012ty}. Finally, we note that $\theta_{13}^\mns = \theta_C / \sqrt{2}$ may also be realised in flavour models without quark-lepton unification, as shown in~\cite{King:2012vj}.}

In this paper we realise these conditions in a specific supersymmetric flavour GUT model.
The model is based on a supersymmetric SU(5) GUT with an $A_4$ family symmetry, supplemented by discrete shaping symmetries and an $R$-symmetry. 
The family symmetry is broken by the vacuum expectation values (VEVs) of flavon fields, which~-- due to an appropriate potential for the flavons~-- point in specific directions of flavour space and lead to the desired patterns of the Yukawa matrices.
Note that, since the Clebsch-Gordan factors play a crucial role for the phenomenological viability of the model, it is essential that we not only provide an effective model valid below the mass scale of messenger fields, but that we also specify the messenger sector.

The model which we construct has, aside from yielding the relation $\theta_{13}^\mns\simeq\theta_C / \sqrt{2}$, several additional interesting aspects:
First, the angle $\alpha=\arg[-V_{td} V_{tb}^*/ (V_{ud} V_{ub}^*)]$ occurring in one of the quark unitarity triangles, which has been measured to be approximately $90^\circ$, results not merely from a fit of the model-parameters to the data, but is a built-in feature in that it is a consequence of the so-called quark phase sum rule~\cite{Antusch:2009hq} which holds whenever  the 1-3 elements of the quark mass matrices vanish.
Second, the CP symmetry is broken only spontaneously due to CP-violating VEVs of the flavon fields.
The spontaneous breaking of CP symmetry allows to greatly reduce the fundamental parameters of the model.
Finally we obtain an excellent fit for the ratio of the strange- to down-quark mass, which is directly related to the ratio of the muon- to electron-mass and a specific combination of the Clebsch factors mentioned above.

The paper is organised as follows.
In section 2, our general strategy for the construction of the flavour model is outlined.
The model is then presented in section~3.
Subsequently, in section~4, we carefully study its phenomenology and predictions, employing a Markov Chain Monte Carlo (MCMC) analysis, before giving the conclusions of the paper in section~5.
The discussion of the superpotential which is responsible for the alignment of the VEVs of the flavon fields, as well as the specification of the renormalizable couplings of the matter-, Higgs-, flavon- and `driving' fields to the messenger fields, is given in the Appendix.

\section{Strategy}
\label{sec:strategy}

Before presenting the model in section~3, let us outline the strategy we followed in order to construct it.
In the following paragraphs we briefly discuss the general features our model shall have.
Afterwards we describe our method to implement these in a concrete model in a consistent way.

\paragraph{General features:}

Recall that in order to obtain $\theta_{13}^\mns \simeq  \theta_C / \sqrt{2}$ we should have $\theta_{12}^d \simeq \theta_C$, as was discussed in Ref.~\cite{Antusch:2012fb}.\footnote{We use the same notation as in Ref.~\cite{Antusch:2009hq}, i.e.\,the moduli of the complex left-mixing angles of $Y_u$, $Y_d$ and $Y_e$ are denoted by $\theta_{ij}^u$, $\theta_{ij}^d$ and $\theta_{ij}^e$, the associated phases by $\delta_{ij}^u$, $\delta_{ij}^d$ and $\delta_{ij}^e$, respectively.}
One attractive possibility to realise the latter is to employ quark Yukawa matrices where the 1-3 (left) mixing angles in the down and up sector each vanish and where thus the 1-3 mixing angle of the Cabibbo-Kobayashi-Maskawa (CKM) mixing matrix is generated from 1-2 and 2-3 rotations.
The 1-2 mixing angles in the Yukawa matrices can then be expressed in terms of CKM angles and the CKM phase as~\cite{Antusch:2009hq}
\be
    \theta^d_{12} \simeq \left|\theta_{12}^\ckm-\frac{\theta_{13}^\ckm}{\theta_{23}^\ckm}e^{-i \delta_\ckm}\right| \simeq 12.0^\circ \pm 0.3^\circ \,, \qquad \theta^u_{12} \simeq \frac{\theta_{13}^\ckm}{\theta_{23}^\ckm} \simeq 5.0^\circ \pm 0.3^\circ \,,
\ee
such that in particular $\theta_{12}^d \simeq \theta_C$ is realised.
Furthermore, under the above assumption of vanishing 1-3 mixings, the quark unitarity triangle angle $\alpha $ is given by the ``quark phase sum rule''~\cite{Antusch:2009hq}
\be
    \alpha \simeq \delta^d_{12}-\delta^u_{12} \;,
\ee
which, when fixed to its experimental value of $\sim 90^\circ$, also implies a realistic CKM CP phase $\delta^\ckm$. 
In our model, we will predict $\delta^d_{12}$, the phase of the 1-2 mixing in the down-quark sector, to be $90^\circ$, via purely real Yukawa matrices except for one single purely imaginary 2-2 entry in $Y_d$.

In this case, SU(5) relations carry over the CP violation into the lepton sector resulting in a Dirac CP phase $\delta^\mns = 90^\circ$ as well. 
Then, the lepton mixing sum rule~\cite{sumrule}
\be
    \theta_{12}^\mns - \theta_{13}^\mns\cot(\theta_{23}^\mns)\cos(\delta^\mns) \simeq \theta_{12}^\nu \;,
\label{eq:sumrule}
\ee
tells us that the PMNS mixing angle $\theta_{12}^\mns$ does not receive significant corrections to the neutrino mixing angle $\theta_{12}^\nu$. 
This calls for tri-bimaximal mixing (TBM) ($\theta_{12}^\nu = 35.3^\circ$, $\theta_{23}^\nu = 45^\circ$, $\theta_{13}^\nu = 0$) in the neutrino sector.
When the neutrino masses are generated via the type I see-saw mechanism~\cite{seesaw}, one simple way to obtain TBM is `constrained sequential dominance'~\cite{King:2002nf}. 
In that scenario, the neutrino masses exhibit a normal mass hierarchy.

In addition to the charged lepton mixing contribution $\theta_{12}^e$, which induces $\theta_{13}^\mns \simeq  \theta_C / \sqrt{2}$ along the lines of~\cite{Antusch:2012fb}, 
our model will also include a charged lepton mixing $\theta_{23}^e$ which can generate a deviation of $\theta_{23}^\mns$ from $45^\circ$, as indicated by recent global fits~\cite{GonzalezGarcia:2012sz,Fogli:2011qn}.

So how can we obtain Yukawa matrices with the features described above?
In our model we generate the rows respectively columns of the Yukawa matrices as VEVs of flavons which are triplets of the family symmetry group $A_4$~\cite{A4}.
In~App.~\ref{sec:flavon} we will provide a superpotential for the flavon fields which gives rise to the VEVs used in our model.

A nice feature of our model is that CP symmetry is broken spontaneously by the flavon VEVs.\footnote{Note that we checked the CP invariance of our superpotential using the `generalised' CP transformation applicable to models with the family group $A_4$, cf.~\cite{Holthausen:2012dk}.}
As was discussed in Ref.~\cite{Antusch:2011sx} (and also briefly in App.~\ref{sec:flavon}), it is in this way possible to obtain Yukawa matrices which are real except for an imaginary 2-2 element in $Y_d$ and $Y_e$.

Concerning the GUT flavour structure of our model, we embed the matter content of the Minimal Supersymmetric Standard Model (MSSM) into representations $\mathbf{\bar 5}$ and $\bf 10$ of SU(5), where the $\mathbf{\bar 5}$ representation 
\be
    F = \begin{pmatrix}d_R^c&d_B^c&d_G^c&e&-\nu\end{pmatrix} \;
\ee
is a triplet under $A_4$, whereas the three $\bf 10$ representations are $A_4$-invariant singlets 
\be
    T_i = \frac{1}{\sqrt{2}}
    \begin{pmatrix}
    0 &-u_G^c & u_B^c & -u_R & -d_R \\
    u_G^c & 0 & -u_R^c & -u_B & -d_B \\
    -u_B^c & u_R^c& 0 & -u_G & -d_G \\
    u_R & u_B& u_G & 0 & -e^c \\
    d_R & d_B& d_G &e^c & 0 \\
    \end{pmatrix}_i\;.
\ee   
Additionally, we introduce, for reasons of minimality, two right-handed neutrinos $N_1$ and $N_2$ as invariant singlets under $A_4$ and SU(5). The SU(5) gauge symmetry is spontaneously broken to the SM gauge group by a VEV of an SU(5) 24-plet $H_{24}$, 
while the VEVs of the fields $H_5$, $H_{\bar 5}$ and $H_{\overline{45}}$ lead to electroweak symmetry breaking.\footnote{We note that while we will explicitly construct the GUT matter sector and the flavour symmetry breaking sector, the details of the (GUT) Higgs sector are beyond the scope of this paper.}

\paragraph{Implementation:}

To implement the desired features described above, we pursue the following procedure:
\begin{itemize}
\item
First, we find a suitable set of effective operators for the matter sector together with an appropriate alignment of flavon VEVs, and a superpotential that generates these VEVs.
\item
The next step is to identify the `shaping symmetry' of the thus specified superpotential. 
The resulting charges for the fields in our model are given in Table~\ref{tab:A4MatterflavSector} on page \pageref{tab:A4MatterflavSector}. 
Note that this result is not unique as one can redefine the resulting $\mathbb{Z}_n$ symmetries by adding/subtracting the columns of charges of the fields in Table~\ref{tab:A4MatterflavSector}.
\item
Next, one has to come up with a set of messenger fields that, when integrated out, can generate all effective operators.
Usually, one has a sizeable freedom of choice in doing so, both from using different representations for the messengers and from using different `topologies' for the diagrams.
For the specific set of messenger fields we choose, see Table~\ref{tab:A4messengerSector} on page \pageref{tab:A4messengerSector}.
\item
Finally, one has to check whether the identified shaping symmetry allows additional effective operators that can spoil the desired features. 
If they are non-renormalizable, one can ignore them if they are not actually generated by the specified set of messengers.
Otherwise one has to modify some choice made in one of the previous steps, be it the alignment, the structure of the effective superpotential or the set of messenger fields.
\end{itemize}

Following this procedure, we developed a model where no such dangerous operators are present and which is thus consistent.

\section{The Model}
\label{sec:model}

We introduce five $A_4$ triplets called $\phi_i$, three $A_4$ invariant singlets called $\xi_i$ and an $A_4$ non-invariant singlet in the $\mathbf{1'}$ representation called $\chi$. They break the family symmetry by their VEVs as given in Table~\ref{tab:flavonvevs}. 
A superpotential which induces these VEVs is presented in App.~\ref{sec:flavon}.
The scale $\Lambda$ is a placeholder for the messenger mass suppression which is specific to that operator which generates the entry (or column respectively row) of the Yukawa matrix (cf. Eqs.~(\ref{YdYeYu}) and~(\ref{yukawa})) where the $\epsilon_i$ appears.
For a full list of the operators, including the messenger fields, which generate the effective superpotential, see the figures on pages~\pageref{fig:diag1}f.
\begin{table}[h]
\centering 
\begin{tabular}{|c|| c| c| c| c| c| c| c| c| c|}
\hline
\rule{0 px}{15 px}
flavon $\varphi_i$:
&$\phi_2$
&$\phi_3$
&$\phi_{ab}$
&$\phi_{N_1}$
&$\phi_{N_2}$
&$\xi_{12}$
&$\xi_{23}$
&$\xi_{M}$ 
&$\chi$\\
\hline
\rule{0 px}{30 px}
$\frac{\vev{\varphi_i}}{\Lambda}$: 
&$\epsilon_2 \begin{pmatrix} 0 \\ 1 \\ 0 \end{pmatrix}$
&$\epsilon_3 \begin{pmatrix} 0 \\ 0 \\ 1 \end{pmatrix} $
&$\epsilon_{ab} \begin{pmatrix} c_{ab} \\ -i s_{ab} \\ 0 \end{pmatrix}$
&$\epsilon_{N_1} \begin{pmatrix} 0 \\  1 \\ -1 \end{pmatrix}$&$ \epsilon_{N_2} \begin{pmatrix} 1 \\  1 \\ 1 \end{pmatrix}$
&$\epsilon_{12}$
&$\epsilon_{23} $
&$\epsilon_{M}$
&$\epsilon_{\chi}$\\[4ex]
\hline
\end{tabular}
\caption{The VEVs of the flavon fields. The $\epsilon_i$ are all assumed to be real numbers. We abbreviated $c_{ab}\equiv \cos (\theta_{ab})$ and $s_{ab}\equiv \sin (\theta_{ab})$.} \label{tab:flavonvevs}
\end{table}

The effective superpotential in the matter sector is given by\footnote{For the sake of brevity, we do not explicitly write down the appropriate suppression by powers of the relevant mass scales $\Lambda$ (which are different for each operator) and we also suppress the couplings in front of each term.}
\begin{equation}
    W_{\text{eff}} = W_{Y_{\nu}} + W_{M_R} + W_{Y_d} + W_{Y_u} \;,
\end{equation}
with
\besub
\label{eq:W_all}
    \begin{align}
        W_{Y_{\nu}} 
        &= (H_5 F) ( N_1\phi_{N_1} + N_2\phi_{N_2} ) \label{eq:wnu}\;,
        \\
        W_{M_R} 
        &= \xi_M^2 ( N_1^2  \phi_{N_1}^2 +  N_2^2  \phi_{N_2}^2 ) \;,
        \\
        W_{Y_d} 
        &= [T_1  H_{\overline{45}}]_{45} \ [F H_{24}]_{\overline{45}}  \ \phi_2 +  [T_2 H_{24}]_{10} \ [F H_{\bar 5}]_{\overline{10}} \ \phi_{ab}
         \nonumber\\
        &  \qquad + [T_3 H_{\bar 5}]_5 \  [F H_{24}]_{\bar{5}}  \ \phi_3  +  [T_3 H_{24}]_{10} \ [F H_{\bar 5}]_{\overline{10}}   \ \chi \phi_2 \;,
        \\
        W_{Y_u} 
        &= H_5 ( T_3^2  +  T_2^2 \phi_{ab}^2+  T_1^2 (\phi_2^2)^2  +  T_2 T_3 \xi_{23} +  T_1 T_2 \xi_{12}^5 ) \;,
    \end{align}
\eesub
where $[X Y]_{R}$ denotes the contraction of the fields $X$ and $Y$ to an SU(5) tensor in the representation $R$.
When $H_{24}$ obtains its VEV and the GUT symmetry gets broken, Clebsch-Gordan factors, which relate the entries of the charged lepton and down-type quark Yukawa matrices, appear.
These factors depend on the way SU(5) indices are contracted and are discrete due to the fact that $H_{24}$ must break SU(5) along the fixed direction of hypercharge. 
In our model, the Clebsch factors $6$, $-1/2$ and $-3/2$,\footnote{More accurately, these are {\it ratios} of Clebsch factors. For a list and analysis of such ratios, see~\cite{Antusch:2009gu}.} arise from contractions mediated by SU(5) 10-plet, $45$-plet and 5-plet messengers coupling to $H_{\bar{5}}$, $H_{\overline{45}}$ and  $H_{\bar{5}}$, respectively, and appear in $Y_e$ (see \eq{YdYeYu}).
Since we need these specific contractions for our model to work, it is essential that we also construct the messenger sector.
The full set of messenger fields, together with their charges, is shown in Table~\ref{tab:A4messengerSector} on page~\pageref{tab:A4messengerSector}.

In order to parametrize the resulting Yukawa matrices, we define the quantities\footnote{Once again we stress that in general specific flavon VEVs do not occur together with a certain messenger field. The suppression of the effective operators by their particular messenger masses and couplings can be obtained from the supergraphs shown in Figures 2--7 on pages~\pageref{fig:diag1}f.}
\be
    \tilde \epsilon_{i} := \frac{\vev{H_{24}}}{\Lambda}\epsilon_{i} \quad \text{and}\quad \hat\epsilon_\chi := \frac{\vev{H_{24}}\vev{\phi_2}}{\Lambda^2} \epsilon_\chi.
\ee

We use the convention of the Particle Data Group~\cite{pdg} for the Yukawa matrices after GUT and flavour symmetry breaking
\be
W = (Y_u^*)_{ij} Q_i u^c_j H_u + (Y^*_d)_{ij} Q_i d^c_j H_d + (Y^*_e)_{ij} L_i e^c_j H_d  + (Y^*_\nu)_{ij} L_i \nu^c_j H_u  + (M_{\nu^c}^*)_{ij} \nu^c_i \nu^c_j\;.
\ee
Up to subleading corrections\footnote{These could, e.g., stem from higher-dimensional operators or from canonical normalisation~\cite{CN}. With the messenger sector of the model specified in the Appendix \ref{app:B} both corrections can be neglected.} 
the Yukawa matrices of the quarks and charged leptons are then given by 
\begin{equation} \label{YdYeYu}
    Y_d = 
    \begin{pmatrix}
        0 & \tilde \epsilon_2 & 0 \\
        \tilde \epsilon_{ab} c_{ab} &  i \tilde \epsilon_{ab} s_{ab} & 0\\
        0 & \omega^2 \hat\epsilon_\chi & \tilde \epsilon_3
    \end{pmatrix},\;
    Y_e = 
    \begin{pmatrix}
        0 & 6 \tilde \epsilon_{ab} c_{ab} & 0 \\
        - \frac 1 2 \tilde \epsilon_2 & i 6 \tilde \epsilon_{ab} s_{ab} & 6 \omega^2 \hat\epsilon_\chi	 \\
        0 & 0& -\frac{3}{2} \tilde \epsilon_3
    \end{pmatrix},\;
    Y_u = \begin{pmatrix}
        \epsilon_2^4 & \epsilon_{12}^5 & 0 \\
        \epsilon_{12}^5 &\epsilon_{ab}^2 & \epsilon_{23} \\
        0 &  \epsilon_{23} &  y_t
    \end{pmatrix},
\end{equation}
where the complex conjugates of the flavon fields build the rows and columns of the down-type quark and charged lepton Yukawa matrices, respectively. The phase $\omega = \exp(2 \pi i/3)$ comes from the fact that the flavon $\chi$ transforms as one-dimensional $\mathbf{1'}$ representation of $A_4$.
The neutrino Yukawa matrix respectively the mass matrix of the heavy neutrinos are given by
\be \label{yukawa}
    Y_\nu = \begin{pmatrix}
    0 &  \epsilon_{N_2} \\
    \epsilon_{N_1} &  \epsilon_{N_2}\\
    -\epsilon_{N_1}  &  \epsilon_{N_2}
    \end{pmatrix}\;,
    \quad
    M_R = 
    \begin{pmatrix} 
    M_{R_1} & 0 \\
    0 & M_{R_2}
    \end{pmatrix} \;.
\ee
The mass matrix for the light neutrinos follows from the see-saw formula~\cite{seesaw}
\be
    m_\nu = \frac{v_u^2}{2}Y_\nu M_R^{-1} Y_\nu^T\;,
\ee
where $v_u = 246 \ \GeV \cdot \sin\beta$.
Inserting \eq{yukawa}, we obtain
\begin{equation}
    m_\nu = \frac{v_u^2}{2}
    \begin{pmatrix}
    A & A & A \\
    A & A + B & A - B\\
    A & A - B & A + B
    \end{pmatrix}\;,\quad\text{with}\quad A =\frac{\epsilon_{N_2}^2}{M_{R_2}}, \;B =\frac{\epsilon_{N_1}^2}{M_{R_1}} \;,
\end{equation}
which implies tri-bimaximal mixing in the neutrino sector.
Note that, since $m_\nu$ depends only on the two parameters $A$ and $B$, we are free to fix two of the four parameters which enter $Y_\nu$ and $M_R$, cf. \eq{Yukawa-pheno}.

\section{Phenomenology}

We presented the Yukawa matrices and the mass matrix of the right-handed neutrinos at the GUT scale $M_{\mathrm{GUT}}= 2\cdot 10^{16}~\GeV$. However, to compare with the experimental data, we need the corresponding low energy values, for instance at the mass scale of the top quark $m_t(m_t)=162.9\ \GeV$. 
Besides the renormalization group (RG) running from $M_{\mathrm{GUT}}$ to $m_t(m_t)$, one needs to include corrections from supersymmetric thresholds when matching the MSSM to the Standard Model at the superpartner mass scale $\Lambda_{\mathrm{SUSY}}$. In this section we discuss the inclusion of these effects into our analysis and present a detailed fit of the parameters at $M_{\mathrm{GUT}}$ to the observables at $m_{t}(m_t)$.

\subsection{Numerical procedure}
\label{subsec:numproc}

The fit is performed in the following way: Using the one-loop MSSM RGEs we run the parameters in the MSSM from $M_\mathrm{GUT}$ down to  $\Lambda_{\mathrm{SUSY}}=1\ \TeV$ with the Mathematica package REAP~\cite{Antusch:2005gp}.  The heavy, right-handed neutrinos are integrated out at their respective mass scales and the effective mass matrix of the light neutrinos is obtained from the see-saw formula. For medium and large $\tan\beta $, SUSY threshold corrections are relevant when matching the MSSM to the Standard Model~\cite{SUSYthresholds,Blazek} at $\Lambda_{\mathrm{SUSY}}$. In our analysis we include the $\tan\beta $-enhanced SUSY threshold corrections in the basis where $Y_u$ is diagonal with the approximate matching relations at $\Lambda_{\mathrm{SUSY}}$
\besub\label{eq:SUSYtreshold}
\begin{align}
Y_d^{\text{SM}} &= (\mathbf{1} + \text{diag} (\eta_{Q_{12}},\eta_{Q_{12}},\eta_{Q_{3}})) \, Y_d^{\text{MSSM}} \cos\beta\;,\\
Y_{u}^{\text{SM}} &= Y_{u}^{\text{MSSM}}\sin\beta\;,\\
Y_e^{\text{SM}} &= Y_e^{\text{MSSM}}\cos\beta\;,
\end{align}
\eesub
where the $\eta_i$ are proportional to $\tan\beta$. We set $\tan\beta=40$ to allow for substantial threshold effects, as required for the Clebsch factors $6$ and $-3/2$ appearing in our model (cf.~\cite{Antusch:2009gu}). The $\eta_i$ parameters can be calculated from the sparticle spectrum. Since we do not specify a certain SUSY scenario, we treat them as free parameters in the fit. In a realistic supersymmetry breaking scenario the SUSY threshold corrections for large $\tan\beta$  typically do not exceed about $50\%$ (see e.g.~\cite{Antusch:2008tf}). In the MCMC analysis we  therefore implement a prior to restrict the SUSY threshold parameters $\eta_{Q_{12}}$ and $\eta_{Q_3}$ to values between $-0.5$ and $0.5$. Note that we have not explicitly included supersymmetric threshold corrections for the charged leptons. They can be absorbed to a good approximation in the quark corrections $\eta_{Q_{12}}$ and $\eta_{Q_{3}}$, since GUTs only predict ratios of quark and charged lepton masses.
Finally we evolve the Yukawa matrices from $\Lambda_{\mathrm{SUSY}}$ to $m_{t}(m_t)$ using the one-loop SM RGEs in REAP, then calculate all observables and compare them to the experimental values. We use the quark- and charged lepton masses at $m_{t}(m_t)$ given in Ref.~\cite{Xing:2007fb} for the fit. Although the masses of the charged leptons are given there to high precision, we set their uncertainty to one percent, which is roughly the accuracy of the one-loop calculation used here. The experimental values of $\sin\theta_C$, $\sin\theta_{23}^\ckm$, $\sin\theta_{13}^\ckm$ and $\delta^\ckm$ are taken from the Winter 2013 fit results of the UTfit collaboration~\cite{UTfit}. The lepton mixing observables are fitted to the updated global fit results of the NuFIT collaboration~\cite{GonzalezGarcia:2012sz}.

There are 14 free parameters in our model which we fit: $\tilde \epsilon_2$, $\tilde \epsilon_3$, $\tilde \epsilon_{ab}$, $\theta_{ab}$, $\hat\epsilon_{\chi}$, $\eta_{Q_{12}}$, $\eta_{Q_{3}}$ for the down-quark and charged-lepton sector and $y_t$, $\epsilon_{ab}$, $\epsilon_2$, $\epsilon_{12}$, $\epsilon_{23}$ for the up-quark sector. In the neutrino sector we choose to fix the masses of the right-handed neutrinos
\be
Y_\nu = \begin{pmatrix}
0 &  \epsilon_{N_2} \\
\epsilon_{N_1} &  \epsilon_{N_2}\\
-\epsilon_{N_1}  &  \epsilon_{N_2}
\end{pmatrix}\;,
\quad
M_R = 
10^{10}~\text{GeV}~\cdot
\begin{pmatrix} 
 1 & 0 \\
0 &   10
\end{pmatrix}\,,
\label{Yukawa-pheno}
\ee
so that the mass matrix of the light neutrinos depends on the parameters  $\epsilon_{N_1}$, $\epsilon_{N_2}$. The choice of the right-handed neutrino masses does not significantly affect the fit as long as we are in the regime where the neutrino Yukawa couplings are $\ll 1$.

We perform a global fit of these 14 parameters to the 18 measured observables (9 fermion masses, 3 quark- and 3 lepton mixing angles, the quark mixing phase and the two neutrino mass-squared differences). Having four more observables than parameters implies that our model is capable of predicting four out of these observables. Therefore, including the yet unknown Dirac CP phase $\delta^\mns$ and the single physical Majorana phase $\varphi_2^\mns$ (using the notation as in REAP~\cite{Antusch:2005gp}) of the PMNS matrix in case of a massless lightest neutrino and normal mass hierarchy, our model makes 6 predictions.

\subsection{Results}

Following the procedure described above, we find a best fit for the parameters with $\chi^2=8.1$. Having 14 parameters and 18 fitted observables this translates to a reduced $\chi^2$ of $\chi^2/\mathrm{d.o.f.}=2.0$. We present the results for the parameters in Table~\ref{tab:pararesults}.\footnote{Note that there is a sign ambiguity for the parameters $\epsilon_{2}$ and $\epsilon_{ab}$, which enter the Yukawa matrix of the up-type quarks at quartic and quadratic order. In our analysis we fix these parameters to be positive.} For a discussion of the hierarchy of the $\epsilon$-parameters we refer to Appendix \ref{app:B}.

The uncertainty of the results is given as highest posterior density (HPD) intervals. These intervals could be interpreted as Bayesian analogues to the frequentist confidence intervals ~\cite{pdg}. The HPD intervals (1$\sigma$ unless stated otherwise) are obtained from a Markov Chain Monte Carlo (MCMC) analysis, using a Metropolis-Hastings algorithm. 
\begin{table}[!h]
\centering 
\begin{tabular}{|rl|r@{.}l|c|}
\hline
\multicolumn{2}{|c|}{Parameter} & \multicolumn{2}{|c|}{Best fit value} & Uncertainty\\
\hline
\hline 
\rule{0 px}{15 px}
$\tilde \epsilon_2$ & in $10^{-4}$ &\hspace{.55cm} $ 6$&$83$ & $_{-0.07}^{+0.10} $ \\
\rule{0 px}{15 px}
$\tilde \epsilon_3$ & in $10^{-1}$& $ 2$&$16$ & $\pm 0.04$\\
\rule{0 px}{15 px}
$\tilde \epsilon_{ab}$ & in $10^{-3}$ & $ -3$&$09$ & $_{-0.04}^{+0.03} $ \\
\rule{0 px}{15 px}
$\theta_{ab}$ & & $ 1$&$319$ & $_{-0.003}^{+0.005} $\\
\rule{0 px}{15 px}
$\hat \epsilon_{\chi}$ & in $10^{-2}$ & $ -1$&$27$ & $_{-0.26}^{+0.21} $ \\
\rule{0 px}{15 px}
$\eta_{Q_{12}}$ & in $10^{-1}$ & $ 3$&$31$ & $_{-3.50}^{+1.45} $ \\ 
\rule{0 px}{15 px}
$\eta_{Q_3}$ & in $10^{-1}$ & $ 1$&$93$ & $_{-0.38}^{+0.49} $ \\ 
\rule{0 px}{15 px}
\multirow{2}{*}{$\epsilon_2$} & 
\multirow{2}{*}{in $10^{-2}$} & 
\multicolumn{2}{|c|}{\multirow{2}{*}{$\begin{cases}5.27\\[2px] 6.07 \end{cases}$}} & 
\multirow{2}{*}{$\begin{cases}_{-0.36}^{+0.34} \\[2px] _{-0.26}^{+0.17} \end{cases}$} \\ 
\rule{0 px}{15 px}
& & \multicolumn{2}{|c|}{} &\\
\rule{0 px}{15 px}
$\epsilon_{ab}$ & in $10^{-2}$ &  $ 4$&$46$ & $_{-0.19}^{+0.75} $\\
\rule{0 px}{15 px}
$\epsilon_{12}$ & in $10^{-1}$  & $ -1$&$65$ & $\pm 0.05$\\
\rule{0 px}{15 px}
$\epsilon_{23}$ & in $10^{-2}$ &  $ 1$&$78$ & $_{-0.22}^{+0.73} $\\
\rule{0 px}{15 px}
$y_t$ & in $10^{-1}$ & $ 5$&$29$ & $_{-0.25}^{+0.29} $\\
\rule{0 px}{15 px}
$\epsilon_{N_1}$ & in $10^{-3}$   & $ 2$&$91$ & $\pm 0.05$\\
\rule{0 px}{15 px}
$\epsilon_{N_2}$ & in $10^{-3}$    & $ 3$&$12$ & $\pm 0.06 $\\[2px]
\hline
\end{tabular}
\caption{Best fit results of the parameters with $\chi^2/\mathrm{d.o.f.}=2.0$. We give $1\sigma$ highest posterior density intervals as uncertainty. The two modes for $\epsilon_2$ can be understood as the two solutions of the (leading order) equation  $y_u\approx\vert(Y_u)_{11}-(Y_u)_{12}^2/(Y_u)_{22}\vert$, where $(Y_u)_{11}=\epsilon_2^4$.} \label{tab:pararesults}
\end{table}

\begin{figure}[p]
\centering
        \begin{subfigure}[b]{0.4\textwidth}
                \centering
                \includegraphics[width=\textwidth]{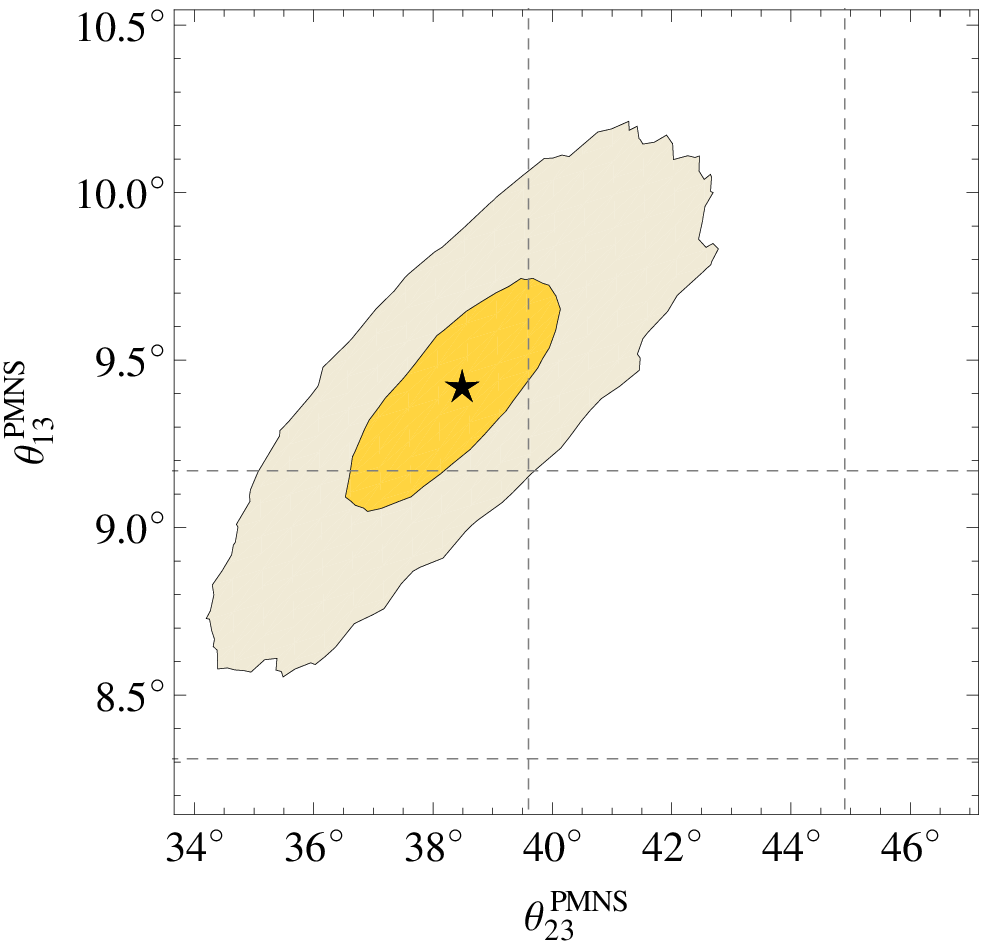}
                \vspace{-.5cm}
                \caption{}
                \label{subfig:t23t13}
        \end{subfigure}
        \quad
        \begin{subfigure}[b]{0.4\textwidth}
                \centering
                \includegraphics[width=\textwidth]{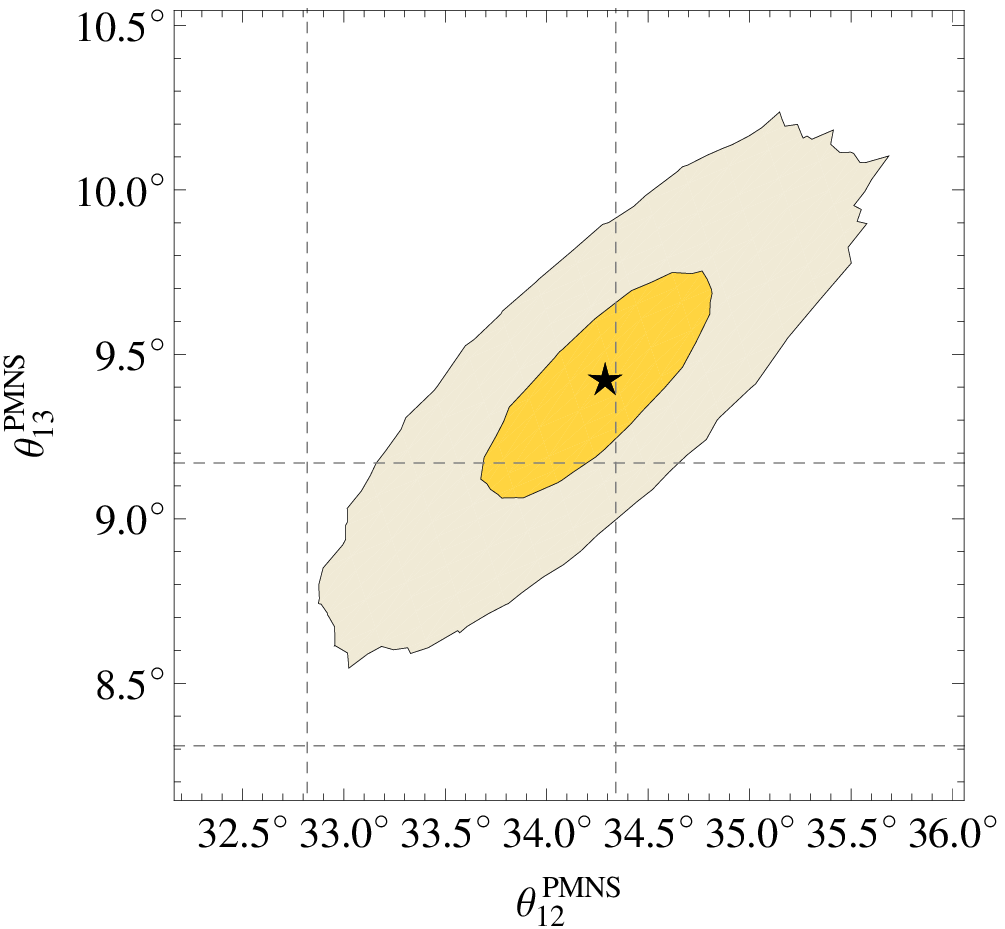}
                \vspace{-.5cm}
                \caption{}
                \label{subfig:t12t13}
        \end{subfigure}
        
        \begin{subfigure}[b]{0.4\textwidth}
                \centering
                \includegraphics[width=\textwidth]{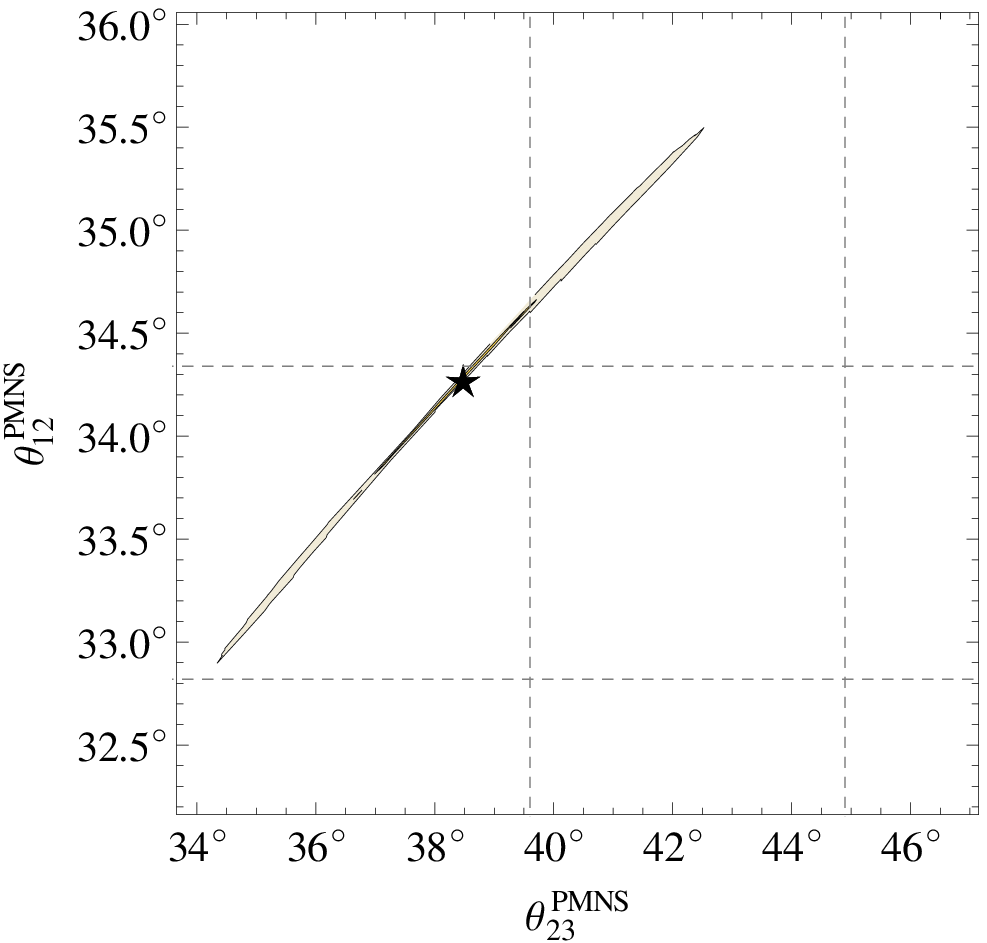}
                \vspace{-.5cm}
                \caption{}
                \label{subfig:t23t12}
        \end{subfigure}
        \quad
        \begin{subfigure}[b]{0.4\textwidth}
                \centering
                \includegraphics[width=\textwidth]{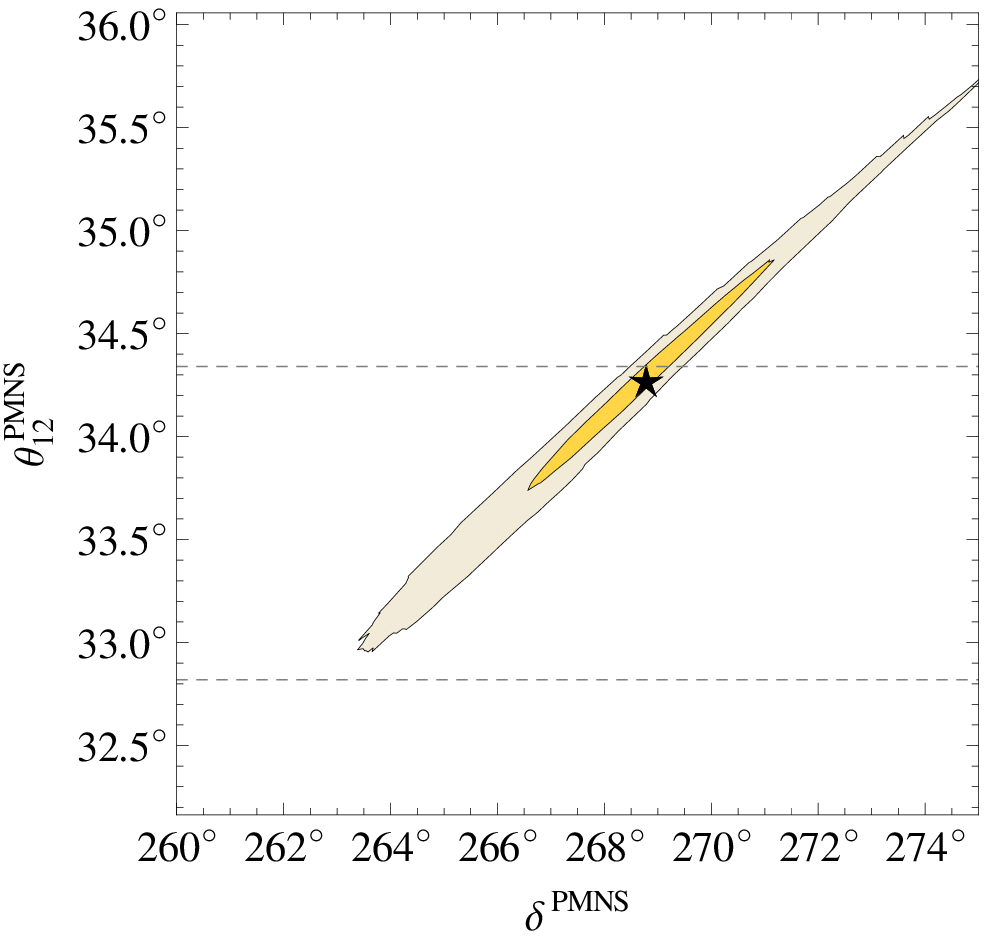}
                \vspace{-.5cm}
                \caption{}
                \label{subfig:dt12}
        \end{subfigure}
        
        \begin{subfigure}[b]{0.4\textwidth}
                \centering          
                \includegraphics[width=\textwidth]{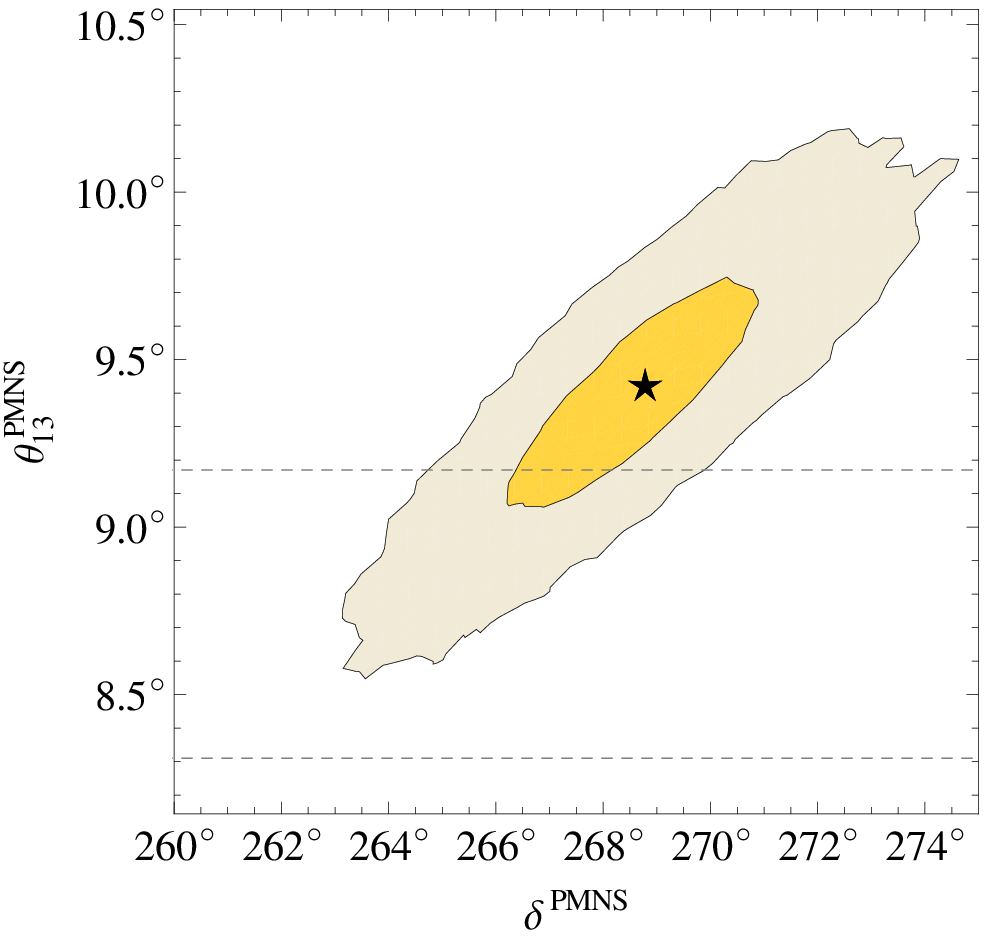}
                \vspace{-.5cm}
                \caption{}
                \label{subfig:dt13}
        \end{subfigure}
		\quad               
        \begin{subfigure}[b]{0.4\textwidth}
                \centering
                \includegraphics[width=\textwidth]{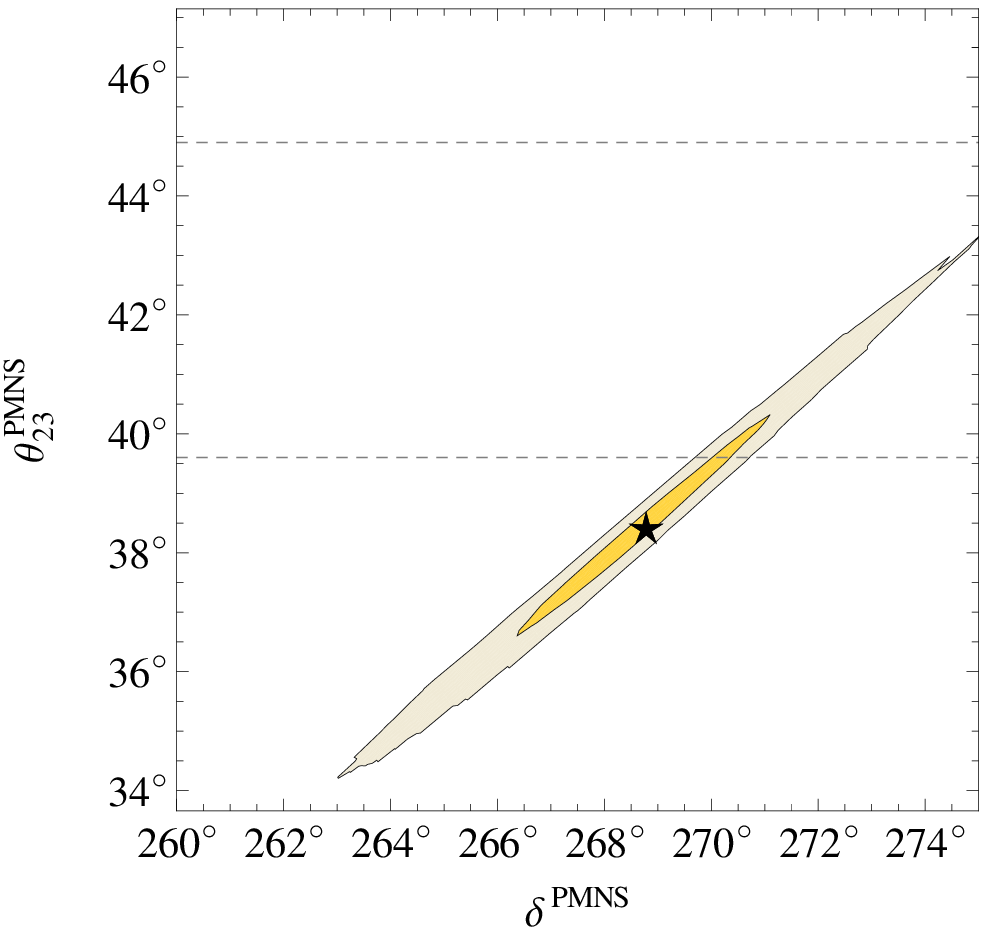}
                \vspace{-.5cm}
                \caption{}
                \label{subfig:dt23}           
        \end{subfigure}
\caption{The plots show the correlations among the lepton mixing angles and the Dirac CP phase. The black star marks the best fit value. The yellow and grey regions give the $1\sigma$ and $3\sigma$ HPD regions obtained from the MCMC analysis, respectively. The dashed grey lines indicate the $1\sigma$ intervals for the measured observables.}
\label{fig:results}
\end{figure}

The corresponding best fit values of the observables at $m_t$ are shown in Table~\ref{tab:results}. Since we used the `$\sin^2$' of the lepton mixing angles as experimental input for the fit, we also present the values of the lepton mixing parameters in degree in Table~\ref{tab:results2}, for convenience. Correlations among the lepton mixing angles and the Dirac CP phase of the MCMC analysis results are plotted in Figure~\ref{fig:results}.
\begin{table}[!h]
\centering 
\begin{tabular}{|rl|rl|r@{.}l|c|}
\hline
\multicolumn{2}{|c|}{Observable} &  \multicolumn{2}{|c|}{Value at $m_{t}$} & \multicolumn{2}{|c|}{Best fit result} & Uncertainty\\
\hline
\hline
\rule{0 px}{15 px}
$m_u$ & in MeV & $1.22$&$_{-0.40}^{+0.48}$ & $1$&$22$ & $ _{-0.40}^{+0.49}$ \\ 
\rule{0 px}{15 px}
$m_c$ & in GeV & $0.59$&$ \pm 0.08$ & $0$&$59$ & $ \pm 0.08$\\
\rule{0 px}{15 px}
$m_t$ & in GeV & $162.9$&$ \pm 2.8$ & $162$&$89$ & $ _{-2.36}^{+2.62}$\\[2px]
\hline
\rule{0 px}{15 px}
$m_d$ & in MeV & $2.76$&$_{-1.14}^{+1.19}$ & $2$&$73$ &  $ _{-0.70}^{+0.30}$\\
\rule{0 px}{15 px}
$m_s$ & in MeV & $52$&$ \pm 15$ & $51$&$66$ &  $_{-13.68}^{+5.60}$\\
\rule{0 px}{15 px}
$m_b$ & in GeV & $2.75$&$ \pm 0.09$ & $2$&$75$ & $ \pm 0.09$\\[2px]
\hline
\rule{0 px}{15 px}
$m_e$ & in MeV & $0.485$&$ \pm 1\%$ & $0$&$483$ & $\pm 0.005$\\
\rule{0 px}{15 px}
$m_\mu$ & in MeV & $102.46$&$ \pm 1\%$ & $102$&$83$ &  $ _{-0.98}^{+1.01}$\\
\rule{0 px}{15 px}
$m_\tau$ & in MeV & $1742$&$ \pm 1\%$ & $1741$&$75$ &  $ _{-17.10}^{+17.38}$\\[2px]
\hline
\rule{0 px}{15 px}
$\sin\theta_C$ & & $0.2254$&$ \pm 0.0007$ & $0$&$2255$ & $ \pm 0.0007$\\
\rule{0 px}{15 px}
$\sin\theta_{23}^\ckm$ & & $0.0421$&$ \pm 0.0006$ & $0$&$0422$ & $ \pm 0.0006$\\
\rule{0 px}{15 px}
$\sin\theta^\ckm_{13}$ && $0.0036$&$ \pm 0.0001$ & $0$&$0036$ & $ \pm 0.0001$\\
\rule{0 px}{15 px}
$\delta^\ckm$ & in $^\circ$& $69.2$&$ \pm 3.1$ & $65$&$65$ &  $ _{-0.53}^{+1.78}$\\[2px]
\hline
\rule{0 px}{15 px}
$\sin^2\theta^\mns_{12}$ & & $0.306$&$ \pm 0.012$ & $0$&$317$ & $ \pm 0.006$\\
\rule{0 px}{15 px}
$\sin^2\theta^\mns_{23}$ && $0.437$&$_{-0.031}^{+0.061}$ & $0$&$387$ & $ _{-0.023}^{+0.017}$\\
\rule{0 px}{15 px}
$\sin^2\theta^\mns_{13}$ && $0.0231$&$ _{-0.0022}^{+0.0023}$ & $0$&$0269$ & $ _{-0.0015}^{+0.0011}$\\
\rule{0 px}{15 px}
$\delta^\mns$ & in $^\circ$& \multicolumn{2}{|c|}{-} & $ 268$&$79$ &  $ _{-1.72}^{+1.32}$\\
\rule{0 px}{15 px}
$\varphi_2^\mns$ & in $^\circ$& \multicolumn{2}{|c|}{-} & $ 297$&$34$ &  $ _{-10.01}^{+8.66}$\\[2px]
\hline
\rule{0 px}{15 px}
$\Delta m^2_{\text{sol}}$ & in $10^{-5}$ eV$^2$ & $7.45$&$ _{-0.16}^{+0.19}$ & $7$&$45$ &  $ _{-0.17}^{+0.18}$\\
\rule{0 px}{15 px}
$\Delta m^2_{\text{atm}}$ & in $10^{-3}$ eV$^2$& $2.421$&$_{-0.023}^{+0.022}$ & $2$&$421$ &  $ _{-0.023}^{+0.022}$\\[2px]
\hline
\end{tabular}
\caption{Best fit results and uncertainties of the observables at $m_t(m_t)$. We give $1\sigma$ highest posterior density intervals as uncertainty. Note that although the masses of the charged leptons are known far more precise than listed here, we set an $1\%$ uncertainty for the experimental values, which is roughly the accuracy of the one loop calculation used here.
\vspace{1cm}
}
\label{tab:results}
\end{table}

\begin{table}[h!]
\centering 
\begin{tabular}{|rl|rl|r@{.}l|c|}
\hline
\multicolumn{2}{|c|}{Observable} & \multicolumn{2}{|c|}{Value at $m_{t}$} & \multicolumn{2}{|c|}{Best fit result} & Uncertainty\\
\hline
\hline
\rule{0 px}{15 px}
$\theta^\mns_{12}$ & in $^\circ$ & $33.57$&$ _{-0.75}^{+0.77}$ &\hspace{.55cm} $34$&$29$ &  $ _{-0.39}^{+0.35}$ \\
\rule{0 px}{15 px}
$\theta^\mns_{23}$ & in $^\circ$ & $41.4$&$_{-1.8}^{+3.5}$ & $38$&$49$ &  $ _{-1.26}^{+1.11}$ \\
\rule{0 px}{15 px}
$\theta^\mns_{13}$ & in $^\circ$& $8.75 $&$ _{-0.44}^{+0.42}$ & $9$&$43$ &  $ _{-0.25}^{+0.20}$ \\[2px]
\hline
\end{tabular}
\caption{Best fit results for the lepton mixing angles at $m_t(m_t)$, given here in degree for convenience.} \label{tab:results2}
\end{table}

\subsubsection{Discussion of the results}
We now discuss how the results shown in Table~\ref{tab:results} and Figure~\ref{fig:results} can be understood from analytic formulae and the Yukawa matrices presented in section 3. 

Let us start with the prediction for the ratio of $m_s$ and $m_d$. From our fit, we obtain
\be\label{eq:msmd}
\frac{m_s}{m_d} = 18.95 _{-0.24}^{+0.33} \;,
\ee
which is in excellent agreement with the value $m_s/m_d \simeq 18.9 \pm 0.8$ obtained from experiments~\cite{Leutwyler:1996qg}. In a small angle approximation of the down-type and charged lepton Yukawa matrices one finds to leading order a simple relation between the ratio of the electron- and muon masses, the ratio of the down- and strange-quark masses and the Clebsch-Gordan coefficients (see e.g.~\cite{Antusch:2011qg})
\be
\frac{m_s}{m_d}\approx\Big\vert\frac{c_{12}c_{21}}{c_{22}^2}\Big\vert\frac{m_\mu}{m_e}\;.
\ee
Although this is just a leading-order estimate, it illustrates well that in order to obtain a valid ratio $m_s/m_d$, a suitable set of Clebsch factors is mandatory. The Clebsch-Gordan factors of our model, $c_{12}=c_{22}=6$ and $c_{21}=-1/2$, are in a remarkably good agreement with the experimental data (cf.\ \eq{eq:msmd}).

For comparison, the often used Clebsch coefficients $c_{12}=c_{21}=1$ and $c_{22}=3$, leading to the Georgi-Jarlskog relations~\cite{GJ}, would result in $m_s/m_d=25.27$, when considering the 1-2 blocks of $Y_e$ and $Y_d$ with zero 1-1 elements and fitting to the experimental values of $\theta_C$, $m_e$ and $m_\mu$. Note that these Clebsch-Gordan coefficients would also not satisfy the equality condition $c_{12}=c_{22}$ of~\cite{Antusch:2012fb} and therefore predict a too small $\theta_{13}^\mns$.

The correlations between the lepton mixing parameters, presented in Figure~\ref{fig:results}, can also be understood from the lepton mixing sum rule
$
\theta_{12}^\mns \simeq\theta_{12}^\nu + \theta_{13}^\mns\cot(\theta_{23}^\mns\vphantom{\frac{a}{a}})\cos(\vphantom{\frac{a}{a}}\delta^\mns)
$
and the relation\footnote{There are two effects leading to a deviation from $\theta_{13}^\mns=\theta_C/\sqrt{2}$. First, the value of $\theta_{23}^\mns<45^\circ$ induces $\sin\theta_{23}^\mns<1/\sqrt{2}$. Second, our model realises $\theta_{12}^e \gtrsim \theta_C$ which results in a correction to larger values of $\theta_{13}^\mns$, as discussed in~\cite{Antusch:2012fb}.} $\theta^\mns_{13}\simeq\theta_C\sin\theta_{23}^\mns$. The latter relation directly explains the correlation between $\theta^\mns_{13}$ and $\theta_{23}^\mns$ in Figure~\ref{subfig:t23t13}. The correlation between $\theta_{23}^\mns$ and $\theta_{12}^\mns$ follows from the lepton mixing sum rule: Larger values of $\theta_{23}^\mns$ have smaller values of $\cot\theta_{23}^\mns$. For more than $90\%$ of the MCMC results $\cos(\delta^\mns)$ is negative. Therefore the values of $\theta_{12}^\mns$ rise with increasing values of $\theta_{23}^\mns$ as can be seen in Figure~\ref{subfig:t23t12}. The correlation between $\theta_{12}^\mns$ and $\theta_{13}^\mns$ is opposite to what one would naively expect from the lepton mixing sum rule. However one also needs to consider the relation $\theta_{13}^\mns \simeq \theta_C\sin\theta^\mns_{23}$, which, when plugged into the lepton mixing sum rule leads to
\be
\theta_{12}^\mns \simeq\theta_{12}^\nu - \theta_C\sqrt{1-(\theta_{13}^\mns)^ 2/\theta_C^2} \ \big\vert\cos(\delta^\mns\vphantom{\frac{a}{a}})\big\vert\;,
\ee
where the negative sign of $\cos\delta^\mns$ is written explicitly. This explains the rising of $\theta^\mns_{12}$ with increasing $\theta^\mns_{13}$ seen in Figure~\ref{subfig:t12t13}. The correlation in Figure~\ref{subfig:dt12} is again obvious from the lepton mixing sum rule. Naively one would not expect correlations between $\delta^\mns$ and the lepton mixing angles $\theta_{13}^\mns$ and $\theta_{23}^\mns$, respectively. The correlations seen in Figures~\ref{subfig:dt13} and~\ref{subfig:dt23} however follow indirectly from the other correlations discussed above.

\subsubsection{Testability of the model}
Finally, we point out predictions which may allow to falsify the model:
\begin{itemize}
\item
We find best fit values of $\theta^\mns_{13} = {9.44^\circ}_{-0.25^\circ}^{+0.20^\circ}$ and $\theta_{23}^\mns={38.49^\circ}_{-1.26^\circ}^{+1.11^\circ}$, which lie within the respective $2\sigma$ intervals reported in \cite{GonzalezGarcia:2012sz}. Future, more precise measurements of $\theta_{13}^\mns$ and $\theta_{23}^\mns$ therefore have the potential to falsify our model.
\item 
For the yet unmeasured Dirac CP phase of the PMNS matrix we predict $\delta^\mns={268.79^\circ}_{-1.72^\circ}^{+1.32^\circ}$.
\item
From the prediction of $\delta^\mns$ follows that the $\cos \delta^\mns$ term in the lepton mixing sum rule leads to a negative correction to the TBM prediction of $35.3^\circ$ for the solar mixing angle. We find $\theta_{12}^\mns ={34.29^\circ} _{-0.40^\circ}^{+0.35^\circ}$.
This is still slightly larger than what is currently reported from experiments.
However, from Figure~\ref{subfig:t23t12} one can deduce that if, as our model predicts, smaller values of $\theta_{23}^\mns$ will be observed, the solar mixing angle will comply with experiments.
\item
In the quark sector, we predict a CKM phase $\delta^\ckm={65.65^\circ}_{-0.53^\circ}^{+1.78^\circ}$. 
This is a slight discrepancy with the experimental value of ${69.2^\circ} \pm 3.1^\circ$, so that a future more precise determination of $\delta^\ckm$ could falsify our model. As already discussed in Eq.~(\ref{eq:msmd}) the model predicts $m_s/m_d= 18.95 _{-0.24}^{+0.33}$, which is currently in excellent agreement with experiments, but will be tested further in the future.
\item
Our prediction for the Majorana phase $\varphi_2^\mns={297.34^\circ} _{-10.01^\circ}^{+8.66^\circ}$ is extremely difficult to be tested, because~-- due to the normal hierarchy of neutrino masses~-- the effective mass parameter which is determined in neutrino-less double beta decay experiments is 
\be
m_{\beta\beta}=\Big\vert (U^\mns_{e2})^2\sqrt{\Delta m^2_{\text{sol}}}+(U^\mns_{e3})^2\sqrt{\Delta m^2_{\text{atm}}}\Big\vert = (2.31 _{-0.09}^{+0.12})\cdot 10^{-3} \, \eV\;,
\ee
which is far below the sensitivity of current experiments. Of course, if an inverse hierarchy would be measured, the model would be falsified.
\end{itemize}

\section{Summary and conclusions}

Motivated by the possibility that the good agreement between $\theta_{13}^\mns$ and $\theta_C/\sqrt{2}$ is no coincidence, we have proposed a first supersymmetric flavour GUT model with SU(5) GUT symmetry and $A_4$ family symmetry, where this relation is realised. Based on one of the strategies discussed in~\cite{Antusch:2012fb}, the neutrino sector of our model features tri-bimaximal mixing, and $\theta_{13}^\mns \simeq  \theta_C / \sqrt{2}$ emerges from the charged lepton contribution to the PMNS matrix, which in turn is linked to quark mixing via specific GUT relations. We explicitly constructed the GUT matter sector of the model, including the full flavon and messenger sectors.

The model has several remarkable properties: For instance, the GUT mixing relations leading to $\theta_{13}^\mns \simeq  \theta_C / \sqrt{2}$ arise after GUT symmetry breaking from a novel combination of group theoretical Clebsch factors, namely $c_{12}=c_{22}=6$ and $c_{21}=-1/2$, which are in excellent agreement with the current experimental data for $m_s/m_d$ (cf.\ \eq{eq:msmd}). 
Furthermore, CP symmetry is broken spontaneously by the VEVs of the flavon fields in a way that the angle $\alpha$ of the quark unitary triangle is close to $90^\circ$ (implying a consistent quark CP phase $\delta^\ckm$). 
This way, also the CP phases of the neutrino sector are predicted, with close-to-maximal CP violation $\delta^\mns \approx 270^\circ$.

Taking into account the RG-evolution of the parameters between the GUT-scale and the electroweak scale, as well as supersymmetric threshold corrections, we have performed a detailed fit of the 14 model parameters to the 18 measured observables. 
We find a good best-fit point with a $\chi^2/$d.o.f.\ of $2.0$. We have also performed a full Markov Chain Monte Carlo fit from which we derive the highest posterior density $1\sigma$ intervals for all parameters and observables. 
With 14 parameters and 18 measured observables plus the PMNS Dirac phase and one Majorana phase, the model features 6 predictions, and  we have discussed how these can be tested by present and future experiments.

\section*{Acknowledgements}
This project was supported by the Swiss National Science Foundation (projects 200021-137513 and CRSII2-141939-1).

\begin{appendix}
\section*{Appendix}

\section{The flavon superpotential} \label{sec:flavon}

Here we present the mechanism responsible for the vacuum alignment of the $A_4$-triplet flavons $\phi_i$, the $A_4$-invariant flavons $\xi_i$ and the flavon $\chi$ in the $\mathbf{1'}$ representation.
The scalar potential for the flavons is obtained from the $F$-term contributions of the `driving fields' $S, D, A$ and $O$ which have $R$-charge $2$. 
Supersymmetric minima occur when the $F$-terms vanish.
All flavons and driving fields are listed in the lower part of Table~\ref{tab:A4MatterflavSector} together with their charges under the imposed symmetries.
Note that the $S$-fields are all singlets with respect to all symmetries (apart from $R$-symmetry) and hence are interchangeable.

The total effective (i.e.~after integrating out the messenger fields) superpotential which is responsible for the vacuum alignment of the flavons, see Table~\ref{tab:flavonvevs}, is given by
\be \label{Wflavon1}
W_{\textrm{flavon}}=W_{\perp}+W_{ab} + W_{N_1 N_2}+W_{\chi}+W_M+ W_{12}
\ee
where\footnote{As we did for the matter-superpotential, we omit the appropriate suppressions by powers of the messenger masses.}$^{,}$\footnote{We use the Ma-Rajasekaran 
 (``$SO(3)$-like'') basis for $A_4$ in this work, as introduced in the first reference of~\cite{A4}. The singlet of $\mathbf{3}\otimes\mathbf{3}$ is given by the $SO(3)$-type inner product. There are two $\mathbf{3}$'s in $\mathbf{3}\otimes\mathbf{3}$, constructed from the antisymmetric cross-product ($\times$) and the symmetric star-product ($\star$), respectively. The brackets $(\dots)_{1^\prime}$ and $(\dots)_{1^{\prime\prime}}$ mean that the fields are contracted to the $\mathbf{1^\prime}$ and $\mathbf{1^{\prime\prime}}$ representation of $A_4$, respectively.} 
\besub \label{Wflavon2}
\bal
W_{\perp}  &= 
S_2 [(\phi_2^{2})^3 - M_2^2] +  S_3 [\phi_3^{2} - M_3^2]   + A_2(\phi_2\star\phi_2)+ A_3(\phi_3\star\phi_3) +O_{2;3}(\phi_2 \phi_3)\,, \\
W_{ab} 
&= S_{ab} [ (\phi_{ab}\star\phi_{ab})^2 \xi_{23}^2  -M_{ab}^2 ]  + D_{ab}^\alpha [ \phi_{ab}^2 + \lambda_{ab} \ \xi_{23}^2 ] + D_{ab}^\beta (\phi_{ab}\star\phi_{ab})\phi_{ab}
\nn \\
&\qquad + D_{ab}^\gamma [ (\phi_{ab}^2)_{1'} (\phi_{ab}^2)_{1''}+ k_{ab} \ (\phi_{ab}\star\phi_{ab})^2 ]
 \,,\\
W_{N_1 N_2}
&= S_{N_1} [ (\phi_{N_1}\star\phi_{N_1})^2  \phi_{N_1}^2 -M_{N_1}^2 ] +  D_{N_1} (\phi_{N_1}\star\phi_{N_1})\phi_{N_1}
+ O_{N_1;N_2} (\phi_{N_1} \phi_{N_2}) \nn\\
&\qquad +S_{N_2} [ (\phi_{N_2}^2)^3-M_{N_2}^2 ] +D'_{N_2} (\phi_{N_2}^2)_{1''}+D''_{N_2} (\phi_{N_2}^2)_{1'}
\,, \\
W_{\chi} &=
S_{\chi}[\chi^6-M_{\chi}^2]
\,,\\
W_{M} &=
S_{M}[\xi_{M}^6-M_{M}^2]
\,,\\
W_{12} &=
S_{12}[\xi_{12}^6-M_{12}^2] \,.
\eal
\eesub
Recall that the constants $M^2$ are real due to the unbroken CP symmetry.
The first terms of each part, which are of the form $S (\varphi^n - M^2)$, restrict the phases of the flavon VEVs to specific discrete values.
In particular, among the discrete vacua for the phases we have $0^\circ$ and $180^\circ$ for $n$ even and $M^2>0$, which we assume everywhere in \eq{Wflavon2}.
Hence, up to a discrete choice, one obtains the real VEVs for $\chi$, $\xi_{M}$ and $\xi_{12}$, see Table~\ref{tab:flavonvevs}.
Let us now briefly discuss the three remaining parts $W_{\perp}$, $W_{ab}$ and $W_{N_1 N_2}$ in the following:
\begin{itemize}
\item[$W_{\perp}$:]
The terms with the triplet driving fields $A_i$ force two of each $\vev{\phi_{i}}$'s components to vanish and the term with the driving field $O_{2;3}$ forces $\vev{\phi_{2}}$ and $\vev{\phi_{3}}$ to be orthogonal. 
Therefore, $W_{\perp}$ has supersymmetric minima at the points $\vev{\phi_2}$ and $\vev{\phi_3}$ specified in Table~\ref{tab:flavonvevs}.
\item[$W_{ab}$:]
The term with the singlet driving field $S_{ab}$ determines the magnitude of the {\it product} of $\vev{\phi_{ab}}\vev{\xi_{23}}$ and its overall phase by the mechanism explained above.
The {\it individual} magnitudes and the relative phase between $\vev{\phi_{ab}}$ and $\vev{\xi_{23}}$ are determined by the \mbox{$F$-term} of $D_{ab}^\alpha$.
The term with the driving field $D_{ab}^\beta$ sets one of the components of $\vev{\phi_{ab}}$ to zero. While the overall phase and norm of $\vev{\phi_{ab}}$ are already fixed by the $F$-terms of $S_{ab}$ and $D_{ab}^\alpha$, the relative magnitude and phase of the two non-vanishing components are fixed by the $F$-term of $D_{ab}^\gamma$. Depending on the value of $k_{ab}$ there are three distinct solutions for these remaining components: For $-1 \leq k_{ab} \leq 3$ the moduli of the two non-vanishing components are equal, while their relative phase $\varphi$ is given by $k_{ab} = 1 - 2 \cos(2 \varphi)$. Defining the ratio of the moduli of the two non-vanishing components as $\tan\theta_{ab}$, the other two solutions are given by $\varphi \in \{0, \pi\}$, $k_{ab} = -1 - 4 \cot^2(2 \theta_{ab})$ for $k_{ab} < -1$ and $\varphi = \pm \frac{\pi}{2}$, $k_{ab} = 3 + 4 \cot^2(2 \theta_{ab})$ for $k_{ab} > 3$, respectively. With an appropriate choice of $k_{ab}$ the potential has a minimum at the points $\vev{\phi_{ab}}$ and $\vev{\xi_{23}}$ given in Table~\ref{tab:flavonvevs}.
\item[$W_{N_1 N_2}$:]
One can easily check that the second line of $W_{N_1 N_2}$ has a supersymmetric minimum at $\vev{\phi_{N_2}} \propto (1,1,1)^T$.
The term with the driving field $D_{N_1}$ in the first line sets one of the components of $\vev{\phi_{N_1}}$ to zero.
Finally, the orthogonality condition between $\vev{\phi_{N_1}}$ and $\vev{\phi_{N_2}}$, which results from setting to zero the $F$-term of $O_{N_1;N_2}$, leads to $\vev{\phi_{N_1}} \propto (0,1,-1)^T$, cf. Table~\ref{tab:flavonvevs}.
\end{itemize}

\section{The messenger sector}
\label{app:B}
In this appendix we elaborate on the set of heavy messenger fields that,  when integrated out, give rise to the effective superpotentials of Section~\ref{sec:model} and Appendix~\ref{sec:flavon}. As discussed in Section~\ref{sec:strategy}, in order not to generate undesired effective operators the messenger sector has to be selected carefully. Note that the choice of the messenger sector is not unique. In the following we present the messenger fields $\Phi_i$ and $\bar \Phi_i$, the supergraphs that give rise to the effective operators and discuss that the numerical values for the $\epsilon$-parameters defined in Eqs.~(\ref{YdYeYu}) and (\ref{yukawa}) can be obtained with a suitable choice for the masses of the messenger fields.

 The quantum numbers of half of the messenger fields $\Phi_i$ are explicitly shown in Table~\ref{tab:A4messengerSector}. For each field $\Phi_i$ there is a  corresponding field $\bar \Phi_i$ with `opposite' quantum numbers, i.e.~quantum numbers such that a mass term $\Lambda_i\Phi_i \bar \Phi_i$ is allowed in the superpotential. 
 
 In Figures 2--7 on pages~\pageref{fig:diag1}f we show the supergraphs that lead to the effective operators when the messenger fields are integrated out. Each supergraph topology in Figures 2--6 corresponds to several operators, which are shown in the table below each graph. In each of the supergraphs, the external fields, which can be either matter-, Higgs-, flavon- or driving-fields, are labeled by $\varphi_i$, whereas the messenger fields are labeled by $\gamma_i$ and $\bar \gamma_i$. Note that many of the messenger fields occur in more than one operator.
 
 Finally let us comment on the consistency of the different orders of magnitude for the numerical values of the $\epsilon$-parameters in Table~\ref{tab:pararesults}.  They arise as effective couplings after integrating out the heavy messenger fields and inserting the vevs for the flavon fields and the GUT-Higgs field in the corresponding non-renormalizable operators. As can be checked from Figures 2--7 on pages \pageref{fig:diag1}f the different orders of magnitude of the parameters, which are needed to fit the data, can originate from an appropriate choice for the masses of the individual messenger fields which appear in the corresponding effective operators. Let us consider the parameters $\tilde\epsilon_2$, $\tilde\epsilon_3$ and $\hat\epsilon_\chi$ as an example. Up to messenger couplings which may be assumed to be $\mathcal{O}(1)$, they are given by
\begin{equation}
\tilde\epsilon_2=\frac{\vev{H_{24}}\vev{\phi_2}}{\Lambda_6\Lambda_7}\,,\quad \tilde\epsilon_3=\frac{\vev{H_{24}}\vev{\phi_3}}{\Lambda_{10}\Lambda_{11}}\,,\quad
\hat\epsilon_\chi=\frac{\vev{H_{24}}\vev{\phi_2}\vev{\chi}}{\Lambda_{32}\Lambda_{33}\Lambda_{34}}\,.
\end{equation}
These messenger mass scales are specific to these parameters and therefore the hierarchy $|\tilde\epsilon_2|\ll|\hat\epsilon_\chi|\ll|\tilde\epsilon_3|$ can arise from a hierarchy of  $\Lambda_{10}\Lambda_{11}\ll\Lambda_{32}\Lambda_{33}\Lambda_{34}/\vev{\chi}\ll\Lambda_6\Lambda_7$. Similar arguments hold for the other parameters.

\end{appendix}


\begin{table}[p]
\centering \small
\begin{tabular}{|c||cc|cc|ccccc|cc|c|}
\hline
\rule{0 px}{15 px}
& SU(5) & $A_4$ & $\mathbb{Z}_2^{(a)}$ & $\mathbb{Z}_2^{(b)}$ & $\mathbb{Z}_6^{(a)}$ & $\mathbb{Z}_6^{(b)}$ & $\mathbb{Z}_6^{(c)}$ & $\mathbb{Z}_6^{(d)}$ & $\mathbb{Z}_6^{(e)}$&$U(1)_a$ & $U(1)_b$ & $U(1)_R$   \\[2ex] 
\hline\hline
\multicolumn{13}{|l|}{\rule{0 px}{15 px}Matter Fields} \\[1ex]
\hline\hline
\rule{0 px}{15 px}
$F$   & 	$\mathbf{\overline{5}}$  &$\mathbf{3}$ 	& \nl & \nl & \nl & 4 & 4 & \nl & \nl & \nl & 2 & 1	\\ 
$T_1$ & 	$\mathbf{10}$ &		\nlone 	& \nl & \nl &\nl & \nl & \nl & \nl & \nl  & \nl & 1 &1	\\
$T_2$ & 	$\mathbf{10}$ 	&		\nlone 	& \nl & \nl & 1 & \nl & 4 & \nl & \nl & \nl & 1 &1	\\
$T_3$ & 	$\mathbf{10}$ & 		\nlone 	& 1  & \nl & \nl & \nl & 2 & \nl &\nl  & \nl & 1 &1	\\
$N_1$ & 	\nlone & 		\nlone  	& \nl & \nl & \nl & 2 & 2 & 1 & \nl & \nl & \nl &1	\\
$N_2$ & 	\nlone & 		\nlone  	& \nl & \nl & \nl & 2 & 2 & 1 & 1   & \nl & \nl &1	
\\[1ex]
\hline\hline
\multicolumn{13}{|l|}{\rule{0 px}{15 px}Higgs Fields} \\ [1ex]
\hline\hline
\rule{0 px}{15 px}
$H_5$ & 			$\mathbf{5}$ & 			\nlone 	& \nl & \nl & \nl & \nl & 2 & \nl  &\nl &\nl & -2  &\nl \\
$H_{\bar 5}$ & 		$\mathbf{\overline{5}}$ & 		\nlone 	& \nl & 1 & \nl & 2 & \nl & \nl  &\nl & -1 & \nl & \nl\\
$H_{45}$ & 		$\mathbf{45}$ & 			\nlone	& \nl & \nl & \nl & 4 & 5 & \nl  & \nl & 1 & \nl & 2 \\
$H_{\overline{45}}$ & 	$\mathbf{\overline{45}}$ & 	\nlone 	& \nl & \nl & \nl & 2 & 1 & \nl  & \nl & -1 & \nl &\nl \\
$H_{24}$     & 		$\mathbf{24}$ & 			\nlone 	& \nl & \nl & \nl & \nl & \nl & \nl  & \nl & 1 & -3 &\nl\\[1ex]   \hline\hline
\multicolumn{13}{|l|}{\rule{0 px}{15 px}Flavon Fields} \\[1ex] 
\hline\hline
\rule{0 px}{15 px}
$\phi_2$     & 		\nlone & $\mathbf{3}$ 	& \nl & \nl & \nl & \nl & 1 &\nl&\nl  & \nl & \nl & \nl \\
$\phi_3$     & 		\nlone & $\mathbf{3}$ 	& 1 & 1 & \nl & \nl & \nl &\nl& \nl & \nl & \nl & \nl\\
$\phi_{ab}$  & 		\nlone & $\mathbf{3}$ 	& \nl & 1 & 5 & \nl & 4 &\nl&\nl  & \nl & \nl & \nl\\ 
$\phi_{N_1}$  & 		\nlone & $\mathbf{3}$ 	& \nl & \nl & \nl & \nl & 4 &5& \nl & \nl & \nl & \nl \\
$\phi_{N_2}$  & 		\nlone & $\mathbf{3}$ 	& \nl & \nl & \nl & \nl & 4 &5& 5 & \nl & \nl & \nl \\
$\chi$ & 			\nlone & $\mathbf{1'}$ 	& 1 & 1 & \nl & \nl & 5 &\nl& \nl & \nl & \nl & \nl \\
$\xi_{12}$  & 		\nlone & \nlone  	& \nl & \nl & 1 & \nl & \nl &\nl& \nl & \nl & \nl & \nl\\
$\xi_{23}$  & 		\nlone & \nlone  	& 1 & \nl & 5 & \nl & 4 &\nl& \nl & \nl & \nl & \nl\\
$\xi_{M}$  & 		\nlone & \nlone  	& \nl & \nl & \nl & 1 & \nl & \nl& \nl & \nl & \nl & \nl\\[1ex]
\hline\hline
\multicolumn{13}{|l|}{\rule{0 px}{15 px}Driving Fields} \\[1ex] 
\hline\hline
\rule{0 px}{15 px}
$S$ & \nlone & \nlone              	& \nl & \nl & \nl & \nl & \nl &\nl& \nl & \nl & \nl & 2  \\
$D_{ab}^\alpha$ & 	\nlone & \nlone 	& \nl & \nl & 2 & \nl & 4 &\nl& \nl & \nl & \nl & 2  \\
$D_{ab}^\beta$ & 	\nlone & \nlone 	& \nl & 1 & 3 & \nl & \nl &\nl&\nl  & \nl & \nl & 2  \\
$D_{ab}^\gamma$ & 	\nlone & \nlone 	& \nl & \nl & 4 & \nl & 2 &\nl& \nl & \nl & \nl & 2  \\
$D_{N_1}$ & 	\nlone & \nlone 	& \nl & \nl & \nl & \nl & \nl &3&\nl  & \nl & \nl & 2  \\
$D'_{N_2}$ & 		\nlone & $\mathbf{1'}$ 	& \nl & \nl & \nl & \nl & 4 &2&2  & \nl & \nl & 2  \\
$D''_{N_2}$ & 		\nlone & $\mathbf{1''}$ 	& \nl & \nl & \nl & \nl & 4 &2&2  & \nl & \nl & 2  \\
$A_2$ & 		\nlone & $\mathbf{3}$    & \nl & \nl & \nl & \nl & 4 &\nl& \nl & \nl & \nl & 2  \\
$A_3$ & 		\nlone & $\mathbf{3}$ 	& \nl & \nl & \nl & \nl & \nl &\nl&\nl  & \nl & \nl & 2  \\
$O_{2;3}$ & 	\nlone & \nlone 	& 1 & 1 & \nl & \nl & 5 &\nl& \nl & \nl & \nl & 2  \\
$O_{N_1;N_2}$& \nlone & \nlone 	& \nl & \nl & \nl & \nl & 4 & 2 & 1 & \nl & \nl & 2  \\[2ex]
 \hline
\end{tabular}
  \caption{The matter-, Higgs-, flavon- and driving fields. A dot means that the field is an invariant singlet under the respective symmetry. Note that the $U(1)$ symmetries will get explicitly broken to $\mathbb{Z}_n$ symmetries by the Higgs sector.}
  \label{tab:A4MatterflavSector} 
\end{table}

\begin{table}[p]
\centering \small
\begin{tabular}{|c||cc|cc|ccccc|cc|c|}
\hline
\rule{0 px}{15 px}
& SU(5) & $A_4$ & $\mathbb{Z}_2^{(a)}$ & $\mathbb{Z}_2^{(b)}$ & $\mathbb{Z}_6^{(a)}$ & $\mathbb{Z}_6^{(b)}$ & $\mathbb{Z}_6^{(c)}$ & $\mathbb{Z}_6^{(d)}$ & $\mathbb{Z}_6^{(e)}$&$U(1)_a$ & $U(1)_b$ & $U(1)_R$     \\[2ex] 
\hline
\hline
\rule{0 px}{15 px}
$\Phi_{1}$ & $\mathbf{\overline{5}}$ & \nlone  & \nl & \nl & \nl & 4 & 2 & 5 & \nl & \nl & 2 & 1  \\
$\Phi_{2}$ & $\mathbf{\overline{5}}$ & \nlone  & \nl & \nl & \nl & 4 & 2 & 5 & 5 & \nl & 2 & 1  \\
$\Phi_{3}$ & $\mathbf{\overline{5}}$ & \nlone  & 1 & \nl &1 & \nl & \nl & \nl & \nl & \nl & 2 & 2  \\
$\Phi_{4}$   & \nlone & $\mathbf{3}$  & \nl & \nl & 2 & \nl & 4 & \nl & \nl & \nl & \nl & 2  \\
$\Phi_{5}$   & \nlone & $\mathbf{3}$  & \nl & \nl & \nl & \nl & 4 & 2 & \nl & \nl & \nl & 2  \\
$\Phi_{6}$   & $\mathbf{\overline{45}}$ & \nlone  & \nl & \nl & \nl & 4 & 5 & \nl & \nl & 1 & -1 & 1 \\
$\Phi_{7}$   & $\mathbf{5}$ & \nlone  & \nl & \nl & \nl & 2 & 1 & \nl & \nl & \nl & -2 & 1  \\
$\Phi_{8}$   & $\mathbf{5}$ & $\mathbf{3}$  & \nl & \nl & 1 & 4 & 2 & \nl & \nl & 1 & \nl & 2  \\
$\Phi_{9}$   & $\mathbf{\overline{10}}$ & \nlone  & \nl & \nl & 5 & \nl & 2 & \nl & \nl & -1 & 2 & 1  \\
$\Phi_{10}$ & $\mathbf{\overline{5}}$ & \nlone  & 1 & 1 & \nl & 4 & 4 & \nl & \nl & 1 & -1 & 1  \\
$\Phi_{11}$ & $\mathbf{5}$ & \nlone  & 1 & 1 & \nl & 2 & 2 & \nl & \nl & \nl & -2 & 1  \\
$\Phi_{12}$   & $\mathbf{\overline{10}}$ & $\mathbf{3}$ & \nl & 1 & \nl & \nl & 4 & \nl & \nl & \nl & -1 & 1  \\
$\Phi_{13}$   & \nlone & $\mathbf{1'}$  & \nl & \nl & 2 & \nl & 4 & \nl & \nl & \nl & \nl & 2  \\
$\Phi_{14}$   & \nlone & $\mathbf{1''}$  & \nl & \nl & 2 & \nl & 4 & \nl & \nl & \nl & \nl & 2  \\
$\Phi_{15}$   & \nlone & \nlone  & \nl & \nl & 4 & \nl & \nl & \nl & \nl & \nl & \nl & 2  \\
$\Phi_{16}$   & \nlone & \nlone  & \nl & \nl & 3 & \nl & \nl & \nl & \nl & \nl & \nl & \nl  \\
$\Phi_{17}$   & \nlone & \nlone  & \nl & \nl & \nl & 4 & \nl & \nl & \nl & \nl & \nl & 2  \\
$\Phi_{18}$   & \nlone & \nlone  & \nl & \nl & \nl & 3 & \nl & \nl & \nl & \nl & \nl & \nl  \\
$\Phi_{19}$   & \nlone & $\mathbf{1'}$  & \nl & \nl & \nl & \nl & 2 & \nl & \nl & \nl & \nl & 2  \\
$\Phi_{20}$   & \nlone & \nlone  & 1 & 1 & \nl & \nl & 3 & \nl & \nl & \nl & \nl & \nl  \\
$\Phi_{21}$   & \nlone & $\mathbf{3}$  & 1 & \nl & 3 & \nl & \nl & \nl & \nl & \nl & \nl & \nl  \\
$\Phi_{22}$ & \nlone & \nlone  & \nl & \nl & \nl & 4 & 4 & 2 & 2 & \nl & \nl & 2  \\
$\Phi_{23}$ & \nlone & \nlone  & \nl & \nl & \nl & \nl & 4 & 2 & 2 & \nl & \nl & 2  \\
$\Phi_{24}$ & \nlone & \nlone  & \nl & \nl & \nl & 4 & 4 & 2 & \nl & \nl & \nl & 2  \\
$\Phi_{25}$ & \nlone & \nlone  & \nl & \nl & \nl & \nl & 4 & 2 & \nl & \nl & \nl & 2  \\
$\Phi_{26}$ & \nlone & \nlone  & \nl & \nl & \nl & \nl & 4 & \nl & \nl & \nl & \nl & 2  \\
$\Phi_{27}$ & \nlone & \nlone  & \nl & \nl & \nl & \nl & 2 & \nl & \nl & \nl & \nl & 2  \\
$\Phi_{28}$ & $\mathbf{10}$ & \nlone  & \nl & \nl & \nl & \nl & 4 & \nl & \nl & \nl & 1 & \nl  \\
$\Phi_{29}$ & \nlone & \nlone  & \nl & \nl & \nl & \nl & 2 & 4 & 4 & \nl & \nl & 2  \\
$\Phi_{30}$ & \nlone & $\mathbf{3}$  & \nl & \nl & \nl & \nl & 2 & 4 & \nl & \nl & \nl & 2  \\
$\Phi_{31}$   & \nlone & \nlone  & \nl & \nl & 5 & \nl & \nl & \nl & \nl & \nl & \nl & \nl  \\
$\Phi_{32}$   & $\mathbf{5}$ & $\mathbf{3}$  & \nl & 1 & \nl & 4 & 5 & \nl & \nl & 1 & \nl & 2  \\
$\Phi_{33}$   & $\mathbf{\overline{10}}$ & $\mathbf{1''}$  & \nl & 1 & \nl & \nl & 5 & \nl & \nl & -1 & 2 & 1  \\
$\Phi_{34}$   & $\mathbf{\overline{10}}$ & $\mathbf{1''}$  & \nl & 1 & \nl & \nl & 5 & \nl & \nl & \nl & -1 & 1  \\[2ex]
\hline
\end{tabular}
\caption{The messenger fields of the model. A dot means that the field is an invariant singlet under the respective symmetry. Note that the $U(1)$ symmetries will get explicitly broken to $\mathbb{Z}_n$ symmetries by the Higgs sector.} 
\label{tab:A4messengerSector}
\end{table}

\newpage

\begin{figure}
\begin{center}
\begin{picture}(200,50)(0,0)
\ArrowLine(0,0)(50,0)
\ArrowLine(100,0)(50,0)
\ArrowLine(100,0)(150,0)
\ArrowLine(200,0)(150,0)
\ArrowLine(50,50)(50,0)
\ArrowLine(150,50)(150,0)
\Text(100,0)[]{$\times$}
\Text(0,10)[]{$\varphi_{1}$}
\Text(45,50)[r]{$\varphi_{2}$}
\Text(75,10)[]{$\gamma_{1}$}
\Text(125,10)[]{$\bar \gamma_{1}$}
\Text(145,50)[r]{$\varphi_{3}$}
\Text(200,10)[]{$\varphi_{4}$}
\end{picture} 

\vspace{0.5cm}
\begin{tabular}{|c || c | c | c | c || c |}\hline\rule{0 px}{15 px}
operator&$\varphi_{1}$&$\varphi_{2}$&$\varphi_{3}$&$\varphi_{4}$&$\gamma_{1}$ \\[1ex]
\hline\rule{0 px}{15 px}
\#1&$N_1$&$H_{5}$&$\phi_{N_1}$&$F$&$\Phi_1$\\
\#2&$N_2$&$H_{5}$&$\phi_{N_2}$&$F$&$\Phi_2$\\
\#3&$\xi_{23}$&$H_{5}$&$T_2$&$T_3$&$\Phi_3$\\
\#4&$\phi_{ab}$&$\phi_{ab}$&$\phi_{ab}$&$D_{ab}^\beta$&$\Phi_4$\\
\#5&$\phi_{N_1}$&$\phi_{N_1}$&$\phi_{N_1}$&$D_{N_1}$&$\Phi_5$\\[1ex]
\hline
\end{tabular}
\caption{List of order 4 operators in the effective superpotential.} \label{fig:diag1}
\end{center}
\end{figure}
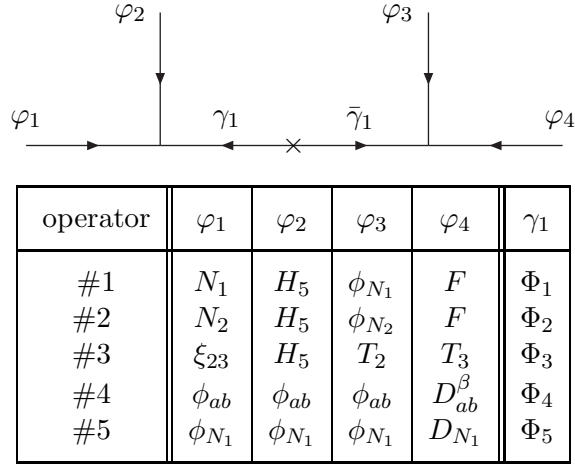
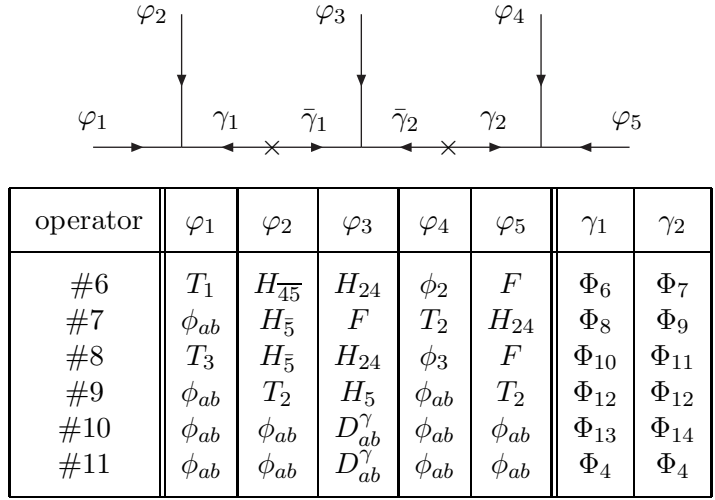
\begin{figure}
\begin{center}
\begin{picture}(200,50)(0,0)
\ArrowLine(0,0)(33,0)
\ArrowLine(67,0)(33,0)
\ArrowLine(67,0)(100,0)
\ArrowLine(133,0)(100,0)
\ArrowLine(133,0)(167,0)
\ArrowLine(200,0)(167,0)
\ArrowLine(33,50)(33,0)
\ArrowLine(100,50)(100,0)
\ArrowLine(167,50)(167,0)
\Text(67,0)[]{$\times$}
\Text(133,0)[]{$\times$}
\Text(0,10)[]{$\varphi_{1}$}
\Text(28,50)[r]{$\varphi_{2}$}
\Text(50,10)[]{$\gamma_{1}$}
\Text(83,10)[]{$\bar \gamma_{1}$}
\Text(95,50)[r]{$\varphi_{3}$}
\Text(117,10)[]{$\bar \gamma_{2}$}
\Text(150,10)[]{$\gamma_{2}$}
\Text(162,50)[r]{$\varphi_{4}$}
\Text(200,10)[]{$\varphi_{5}$}
\end{picture} 

\vspace{0.5cm}
\begin{tabular}{|c || c | c | c | c | c || c | c |} \hline\rule{0 px}{15 px}
operator&$\varphi_{1}$&$\varphi_{2}$&$\varphi_{3}$&$\varphi_{4}$&$\varphi_{5}$&$\gamma_{1}$&$\gamma_{2}$ \\[1ex]
\hline\rule{0 px}{15 px}
\#6&$T_1$&$H_{\overline{45}}$&$H_{24}$&$\phi_2$&$F$&$\Phi_6$&$\Phi_7$\\
\#7&$\phi_{ab}$&$H_{\bar 5}$&$F$&$T_2$&$H_{24}$&$\Phi_8$&$\Phi_{9}$\\
\#8&$T_3$&$H_{\bar 5}$&$H_{24}$&$\phi_3$&$F$&$\Phi_{10}$&$\Phi_{11}$\\
\#9&$\phi_{ab}$&$T_2$&$H_{ 5}$&$\phi_{ab}$&$T_2$&$\Phi_{12}$&$\Phi_{12}$\\
\#10&$\phi_{ab}$&$\phi_{ab}$&$D_{ab}^\gamma$&$\phi_{ab}$&$\phi_{ab}$&$\Phi_{13}$&$\Phi_{14}$\\
\#11&$\phi_{ab}$&$\phi_{ab}$&$D_{ab}^\gamma$&$\phi_{ab}$&$\phi_{ab}$&$\Phi_{4}$&$\Phi_{4}$\\[1ex]
\hline
\end{tabular}
\caption{List of order 5 operators in the effective superpotential.} \label{fig:diag2}
\end{center}
\end{figure}
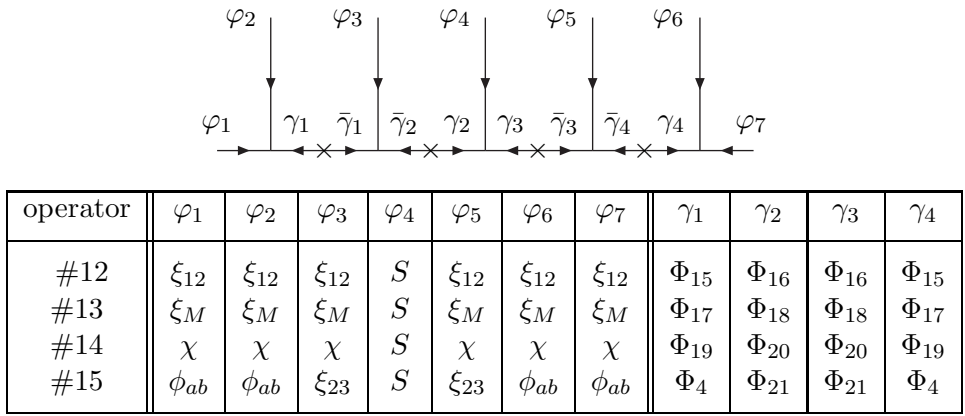
\begin{figure}
\begin{center}
\begin{picture}(200,50)(0,0)
\ArrowLine(0,0)(20,0)
\ArrowLine(40,0)(20,0)
\ArrowLine(40,0)(60,0)
\ArrowLine(80,0)(60,0)
\ArrowLine(80,0)(100,0)
\ArrowLine(120,0)(100,0)
\ArrowLine(120,0)(140,0)
\ArrowLine(160,0)(140,0)
\ArrowLine(160,0)(180,0)
\ArrowLine(200,0)(180,0)
\ArrowLine(20,50)(20,0)
\ArrowLine(60,50)(60,0)
\ArrowLine(100,50)(100,0)
\ArrowLine(140,50)(140,0)
\ArrowLine(180,50)(180,0)
\Text(40,0)[]{$\times$}
\Text(80,0)[]{$\times$}
\Text(120,0)[]{$\times$}
\Text(160,0)[]{$\times$}
\Text(0,10)[]{$\varphi_{1}$}
\Text(15,50)[r]{$\varphi_{2}$}
\Text(55,50)[r]{$\varphi_{3}$}
\Text(95,50)[r]{$\varphi_{4}$}
\Text(135,50)[r]{$\varphi_{5}$}
\Text(175,50)[r]{$\varphi_{6}$}
\Text(200,10)[]{$\varphi_{7}$}
\Text(30,10)[]{$\gamma_{1}$}
\Text(50,10)[]{$\bar \gamma_{1}$}
\Text(70,10)[]{$\bar \gamma_{2}$}
\Text(90,10)[]{$\gamma_{2}$}
\Text(110,10)[]{$\gamma_{3}$}
\Text(130,10)[]{$\bar \gamma_{3}$}
\Text(150,10)[]{$\bar \gamma_{4}$}
\Text(170,10)[]{$\gamma_{4}$}
\end{picture} 

\vspace{0.5cm}
\begin{tabular}{|c || c | c | c | c | c | c | c || c | c | c | c |} \hline
operator&$\varphi_{1}$&$\varphi_{2}$&$\varphi_{3}$&$\varphi_{4}$&$\varphi_{5}$&$\varphi_{6}$&$\varphi_{7}$&$\gamma_{1}$&$\gamma_{2}$&$\gamma_{3}$&$\gamma_{4}$ \\[1ex]
\hline\rule{0 px}{15 px}
\#12&$\xi_{12}$&$\xi_{12}$&$\xi_{12}$&$S$&$\xi_{12}$&$\xi_{12}$&$\xi_{12}$&$\Phi_{15}$&$\Phi_{16}$&$\Phi_{16}$&$\Phi_{15}$\\
\#13&$\xi_M$&$\xi_M$&$\xi_M$&$S$&$\xi_M$&$\xi_M$&$\xi_M$&$\Phi_{17}$&$\Phi_{18}$&$\Phi_{18}$&$\Phi_{17}$\\
\#14&$\chi$&$\chi$&$\chi$&$S$&$\chi$&$\chi$&$\chi$&$\Phi_{19}$&$\Phi_{20}$&$\Phi_{20}$&$\Phi_{19}$\\
\#15&$\phi_{ab}$&$\phi_{ab}$&$\xi_{23}$&$S$&$\xi_{23}$&$\phi_{ab}$&$\phi_{ab}$&$\Phi_4$&$\Phi_{21}$&$\Phi_{21}$&$\Phi_4$\\[1ex]
\hline
\end{tabular}
\caption{List of order 7 operators in the effective superpotential from supergraphs with linear topology.} \label{fig:diag4}
\end{center}
\end{figure}
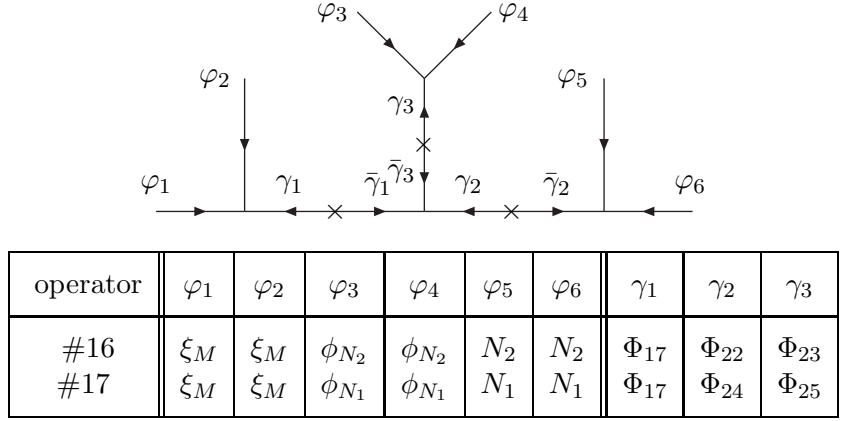
\begin{figure}
\begin{center}
\begin{picture}(200,75)(0,0)
\ArrowLine(0,0)(33,0)
\ArrowLine(67,0)(33,0)
\ArrowLine(67,0)(100,0)
\ArrowLine(133,0)(100,0)
\ArrowLine(133,0)(167,0)
\ArrowLine(200,0)(167,0)
\ArrowLine(33,50)(33,0)
\ArrowLine(75,75)(100,50)
\ArrowLine(100,25)(100,50)
\ArrowLine(100,25)(100,0)
\ArrowLine(125,75)(100,50)
\ArrowLine(167,50)(167,0)
\Text(67,0)[]{$\times$}
\Text(133,0)[]{$\times$}
\Text(100,25)[]{$\times$}
\Text(0,10)[]{$\varphi_{1}$}
\Text(28,50)[r]{$\varphi_{2}$}
\Text(72,75)[r]{$\varphi_{3}$}
\Text(128,75)[l]{$\varphi_{4}$}
\Text(162,50)[r]{$\varphi_{5}$}
\Text(200,10)[]{$\varphi_{6}$}
\Text(50,10)[]{$\gamma_{1}$}
\Text(83,10)[]{$\bar \gamma_{1}$}
\Text(117,10)[]{$\gamma_{2}$}
\Text(150,10)[]{$\bar  \gamma_{2}$}
\Text(97,40)[r]{$\gamma_{3}$}
\Text(97,15)[r]{$\bar  \gamma_{3}$}
\end{picture} 

\vspace{0.5cm}
\begin{tabular}{|c || c | c | c | c | c | c || c | c | c |}\hline\rule{0 px}{15 px}
operator&$\varphi_{1}$&$\varphi_{2}$&$\varphi_{3}$&$\varphi_{4}$&$\varphi_{5}$&$\varphi_{6}$&$\gamma_{1}$&$\gamma_{2}$&$\gamma_{3}$ \\[1ex]
\hline\rule{0 px}{15 px}
\#16&$\xi_M$&$\xi_M$&$\phi_{N_2}$&$\phi_{N_2}$&$N_2$&$N_2$&$\Phi_{17}$&$\Phi_{22}$&$\Phi_{23}$\\
\#17&$\xi_M$&$\xi_M$&$\phi_{N_1}$&$\phi_{N_1}$&$N_1$&$N_1$&$\Phi_{17}$&$\Phi_{24}$&$\Phi_{25}$\\[1ex]
\hline
\end{tabular}
\caption{List of order 6 operators in the effective superpotential from supergraphs with non-linear topology.} \label{fig:diag5}
\end{center}
\end{figure}
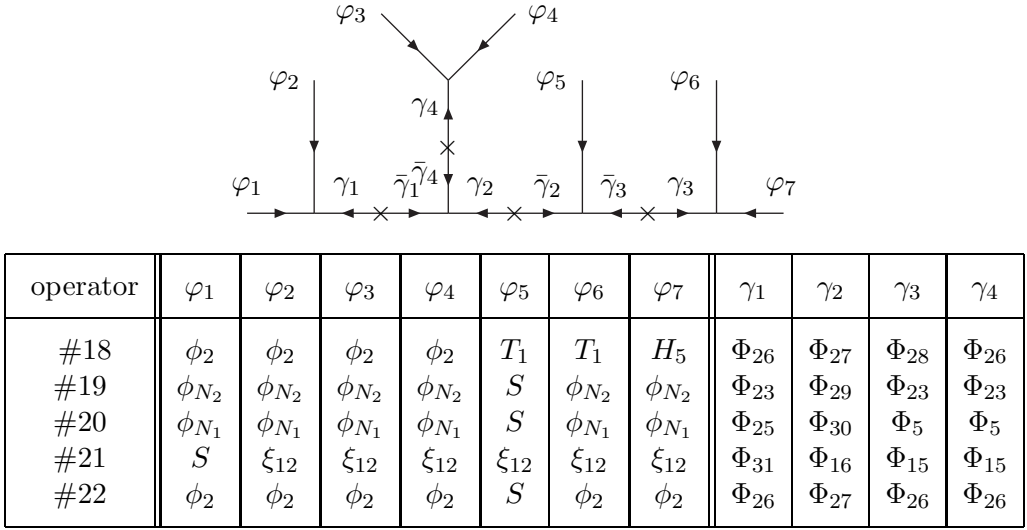
\begin{figure}
\begin{center}
\begin{picture}(200,75)(0,0)
\ArrowLine(0,0)(25,0)
\ArrowLine(50,0)(25,0)
\ArrowLine(50,0)(75,0)
\ArrowLine(100,0)(75,0)
\ArrowLine(100,0)(125,0)
\ArrowLine(150,0)(125,0)
\ArrowLine(150,0)(175,0)
\ArrowLine(200,0)(175,0)
\ArrowLine(25,50)(25,0)
\ArrowLine(50,75)(75,50)
\ArrowLine(75,25)(75,0)
\ArrowLine(75,25)(75,50)
\ArrowLine(100,75)(75,50)
\ArrowLine(125,50)(125,0)
\ArrowLine(175,50)(175,0)
\Text(50,0)[]{$\times$}
\Text(100,0)[]{$\times$}
\Text(150,0)[]{$\times$}
\Text(75,25)[]{$\times$}
\Text(0,10)[]{$\varphi_{1}$}
\Text(20,50)[r]{$\varphi_{2}$}
\Text(45,75)[r]{$\varphi_{3}$}
\Text(105,75)[l]{$\varphi_{4}$}
\Text(120,50)[r]{$\varphi_{5}$}
\Text(170,50)[r]{$\varphi_{6}$}
\Text(200,10)[]{$\varphi_{7}$}
\Text(37.5,10)[]{$\gamma_{1}$}
\Text(60,10)[]{$\bar \gamma_{1}$}
\Text(87.5,10)[]{$\gamma_{2}$}
\Text(112.5,10)[]{$\bar \gamma_{2}$}
\Text(137.5,10)[]{$\bar \gamma_{3}$}
\Text(162.5,10)[]{$ \gamma_{3}$}
\Text(72,40)[r]{$\gamma_{4}$}
\Text(72,15)[r]{$\bar  \gamma_{4}$}
\end{picture} 

\vspace{0.5cm}
\begin{tabular}{|c || c | c | c | c | c | c | c || c | c | c | c |} \hline\rule{0 px}{15 px}
operator&$\varphi_{1}$&$\varphi_{2}$&$\varphi_{3}$&$\varphi_{4}$&$\varphi_{5}$&$\varphi_{6}$&$\varphi_{7}$&$\gamma_{1}$&$\gamma_{2}$&$\gamma_{3}$&$\gamma_{4}$ \\[1ex]
\hline\rule{0 px}{15 px}
\#18&$\phi_{2}$&$\phi_{2}$&$\phi_{2}$&$\phi_{2}$&$T_1$&$T_1$&$H_5$&$\Phi_{26}$&$\Phi_{27}$&$\Phi_{28}$&$\Phi_{26}$\\
\#19&$\phi_{N_2}$&$\phi_{N_2}$&$\phi_{N_2}$&$\phi_{N_2}$&$S$&$\phi_{N_2}$&$\phi_{N_2}$&$\Phi_{23}$&$\Phi_{29}$&$\Phi_{23}$&$\Phi_{23}$\\
\#20&$\phi_{N_1}$&$\phi_{N_1}$&$\phi_{N_1}$&$\phi_{N_1}$&$S$&$\phi_{N_1}$&$\phi_{N_1}$&$\Phi_{25}$&$\Phi_{30}$&$\Phi_{5}$&$\Phi_{5}$\\
\#21&$S$&$\xi_{12}$&$\xi_{12}$&$\xi_{12}$&$\xi_{12}$&$\xi_{12}$&$\xi_{12}$&$\Phi_{31}$&$\Phi_{16}$&$\Phi_{15}$&$\Phi_{15}$\\
\#22&$\phi_{2}$&$\phi_{2}$&$\phi_{2}$&$\phi_{2}$&$S$&$\phi_{2}$&$\phi_{2}$&$\Phi_{26}$&$\Phi_{27}$&$\Phi_{26}$&$\Phi_{26}$\\[1ex]
\hline
\end{tabular}
\caption{List of order 7 operators in the effective superpotential from supergraphs with non-linear topology.} \label{fig:diag6}
\end{center}
\end{figure}
\begin{figure}
\noindent
\begin{picture}(200,50)(0,0)
\ArrowLine(0,0)(25,0)
\ArrowLine(50,0)(25,0)
\ArrowLine(50,0)(75,0)
\ArrowLine(100,0)(75,0)
\ArrowLine(100,0)(125,0)
\ArrowLine(150,0)(125,0)
\ArrowLine(150,0)(175,0)
\ArrowLine(200,0)(175,0)
\ArrowLine(25,50)(25,0)
\ArrowLine(75,50)(75,0)
\ArrowLine(125,50)(125,0)
\ArrowLine(175,50)(175,0)
\Text(50,0)[]{$\times$}
\Text(100,0)[]{$\times$}
\Text(150,0)[]{$\times$}
\Text(0,10)[]{$\phi_2$}
\Text(20,50)[r]{$H_{\bar 5}$}
\Text(70,50)[r]{$F$}
\Text(120,50)[r]{$H_{24}$}
\Text(170,50)[r]{$T_3$}
\Text(200,10)[]{$\chi$}
\Text(37.5,10)[]{$\Phi_{32}$}
\Text(62.5,10)[]{$\bar \Phi_{32}$}
\Text(87.5,10)[]{$\bar \Phi_{33}$}
\Text(112.5,10)[]{$\Phi_{33}$}
\Text(137.5,10)[]{$\bar \Phi_{34}$}
\Text(162.5,10)[]{$\Phi_{34}$}
\end{picture} 
\hspace{1cm}
\begin{picture}(200,75)(0,0)
\ArrowLine(0,0)(20,0)
\ArrowLine(40,0)(20,0)
\ArrowLine(40,0)(60,0)
\ArrowLine(80,0)(60,0)
\ArrowLine(80,0)(100,0)
\ArrowLine(120,0)(100,0)
\ArrowLine(120,0)(140,0)
\ArrowLine(160,0)(140,0)
\ArrowLine(160,0)(180,0)
\ArrowLine(200,0)(180,0)
\ArrowLine(20,50)(20,0)
\ArrowLine(60,50)(60,0)
\ArrowLine(80,75)(100,50)
\ArrowLine(100,25)(100,0)
\ArrowLine(100,25)(100,50)
\ArrowLine(120,75)(100,50)
\ArrowLine(140,50)(140,0)
\ArrowLine(180,50)(180,0)
\Text(40,0)[]{$\times$}
\Text(80,0)[]{$\times$}
\Text(100,25)[]{$\times$}
\Text(120,0)[]{$\times$}
\Text(160,0)[]{$\times$}
\Text(0,10)[]{$\xi_{12}$}
\Text(15,50)[r]{$\xi_{12}$}
\Text(55,50)[r]{$\xi_{12}$}
\Text(75,75)[r]{$\xi_{12}$}
\Text(125,75)[l]{$\xi_{12}$}
\Text(135,50)[r]{$T_2$}
\Text(175,50)[r]{$T_1$}
\Text(200,10)[]{$H_5$}
\Text(30,-10)[]{$\Phi_{15}$}
\Text(50,-10)[]{$\bar \Phi_{15}$}
\Text(70,-10)[]{$\bar \Phi_{16}$}
\Text(90,-10)[]{$\Phi_{16}$}
\Text(110,-10)[]{$\bar  \Phi_{31}$}
\Text(130,-10)[]{$\Phi_{31}$}
\Text(150,-10)[]{$\bar  \Phi_{28}$}
\Text(170,-10)[]{$\Phi_{28}$}
\Text(97,40)[r]{$\Phi_{15}$}
\Text(97,15)[r]{$\bar \Phi_{15}$}
\end{picture}
\vspace{.5cm}
\caption{Additional effective operators of order 6: $[T_3 H_{24}]_{10} \ [F H_{\bar 5}]_{\overline{10}}   \ \chi \phi_2 $ (left) and of order 8: $T_1 T_2 \xi_{12}^5 H_5$ (right).} \label{fig:diag37}
\end{figure}
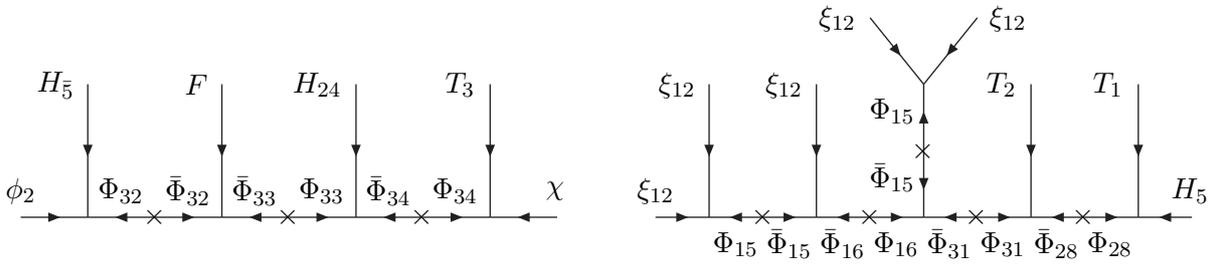

\end{document}